# QUANTUM ZENO EFFECT, KAPITSA PENDULUM AND SPINNING TOP PRINCIPLE.  COMPARATIVE ANALYSIS

**V.A. Buts**
National Science Center «Kharkiv Institute of Physics and Technology»
Akademicheskaya str. 1, 61108 Kharkiv;  V.N. Karazin Kharkiv National University
Svobody Sq. 4, 61022, Kharkiv;     Institute of Radio Astronomy of NAS of Ukraine
4 Chervonopraporna Str. Kharkov 61002, Ukraine
e-mail: vbuts@kipt.kharkov.ua

Comparative analysis of three stabilization mechanisms of unstable states of physical systems is presented in this review. These mechanisms are: the quantum Zeno effect, the stabilization of unstable states in an external fast oscillating field (i.e. the Kapitza pendulum), and the mechanism named as the spinning top mechanism. The common features of these mechanisms, as well as the differences between them, are analyzed in the paper. In particular, it is shown that the stabilization of quantum systems is possible without involvement of such concept as the collapse of wave function. For stabilization it is sufficient to have a stabilizing radiation flow with the Rabi frequency of transitions exceeding some frequency. This frequency is inversely proportional to the lifetime of the state under stabilization. It is shown that the Top principle allows stabilizing unstable systems using affecting only those states to which these systems must go over.
    It is shown that stabilization of unstable states by impact of rapidly oscillating forces occurs by non-self-consistent exposure, i.e. the dynamics of stabilizing field is independent on the dynamics of the stabilized state. Stabilization using the spinning top principle involves self-consistent forces, and thus, in many cases can be the most effective mechanism of stabilization.

The quantum Zeno effect and the mechanism of a pendulum stabilization using an inverted suspension (the Kapitsa pendulum) are the two well-known mechanisms that, when created conditions for their realization, stabilize initially unstable stationary states of many physical systems. A common characteristic of these mechanisms is the rapid, periodic influence acting upon the unstable system. For the quantum Zeno effect, this influence consists of the frequent observation of the unstable quantum system. The instability of a vertical position of a mathematical pendulum can be suppressed by a rapid change in the position of the suspension of the pendulum. These are well-known examples of stabilization of unstable quantum and classical systems. Another way to stabilize an unstable system is to involve it in some additional rapid periodic motion. The properties of this motion necessary to transform an unstable system into a stable system are well known to anybody who seen a simple and obvious physical prototype of this stabilization – a children's toy: spinning top.  The vertical position of a top without a rotation is unstable. The time of the spinning top fall (the lifetime of the vertical position of the spinning top) can be attributed to the lifetime of the unstable state. If the spinning top is rotated and the period of its rotation is much shorter than the lifetime of the unstable state, the vertical position is stable. By analogy to this process, the way of suppressing instabilities using rapid periodic motion can be named the spinning top principle (mechanism). It can be seen that all three mechanisms mentioned have a common feature - a rapid periodic change of the characteristics of the systems being stabilized. There are, however, the differences. Here we analyze such differences.

First of all, we consider the quantum Zeno effect. Further, using the example of Kapitsa's pendulum, we compare the spinning top principle with the stabilization mechanism using an external rapidly oscillating field. The difference between these mechanisms appears to be the fact that in a spinning top type of system the stabilizing disturbance is self-consistent with the dynamics of the system being stabilized. Stabilizing perturbation in a pendulum with an inverted suspension is independent on the pendulum. It is shown that in those cases when the spinning top principle can be used, this stabilization proves to be more effective than the mechanism of stabilization by an external rapidly oscillating field. A brief description of examples to use the spinning top principle for the stabilization of various physical systems is provided in the fourth section. In conclusion, the main results of the work are discussed.

### STABILIZATION OF QUANTUM SYSTEMS. QUANTUM ZENO EFFECT.

The extensive literature is devoted to the quantum effect of Zeno. A simple internet search brings both a description of the effect itself and the description of the latest experimental observations of this effect. Simplest way it is to look at https://en.wikipedia.org/wiki/Quantum_Zeno_effect and at their bibliography. Here we only note that the roots of Zeno effect lie in two fundamental concepts: 1. Non-exponential law of decay of quantum excited states at small time intervals and 2.The collapse of the wave function in the process of measurement. Below we draw attention to the fact that the first fundamental feature of quantum systems was described as early as 1947 by V.A. Fock and N.S. Krylov [1]. The second concept - the measurement process (and, correspondingly, the collapse of the wave functions) cannot be described in the framework of traditional quantum mechanics. However, the measurement process is not necessary for stabilization of unstable states, as it is shown below. It suffices to account for some external perturbation which is fast enough for periodically transfer the system back and forward from one state to another. Such process is fully described in the framework of traditional quantum mechanics. Therefore, we do not consider the process of measurement (the collapse of the wave function) below. It should be noted that often the process of external influence, which transforms the system to another state, is identified with the measurement process.

In [1] authors obtain a general universal expression for the time dependence of the probability of finding quantum system in the initial state $L(t)$. This expression is often known as the Krylov-Fock theorem and has the form:

$$L(t) = \left| \int w(E) \cdot \exp(-i \cdot E \cdot t / \hbar) \cdot dE \right|^2, \qquad (1)$$

where $w(E)$ - is the differential distribution function of the initial state (density of the distribution function); $w(E) \cdot dE$ - energy spectrum of the initial state.

The authors carried out a rather detailed analysis of this expression. We note some of the results that follow from the analysis of expression (1). First of all, the expression under the modulus in (1) is a characteristic function. Therefore, it is limited. Moreover, in our case it is less or equal than one ( $L(t) = |p(t)|^2 \leq 1$ ). If the probability density is an absolutely integrable function, then it is easy to see that the probability at infinity tends to zero ($\lim_{t\to\infty} L(t) \to 0$). The most interesting feature of this function is the behavior at small times. The function $L(t)$, like any characteristic function, can be represented in the form of an expansion in a series of moments:

$$L(t) = \left| \sum_{n=0}^{\infty} \left( \frac{i \cdot t}{\hbar} \right)^n \cdot \frac{M_n}{n!} \right|^2 . \tag{2}$$

here $M_n \equiv \langle E^n \rangle = \int_{-\infty}^{\infty} E^n w(E) dE$.

Restricting ourselves to the first three terms in the sum at small time intervals, we obtain:

$$L(t) = 1 - \left( \frac{t}{\hbar} \right)^2 \langle (\Delta E)^2 \rangle , \tag{3}$$

where $\Delta E \equiv E - E_0$, $M_1 \equiv \langle E \rangle \equiv E_0 = \int_{-\infty}^{\infty} E w(E) dE.$

This expression coincides with the one obtained in [1].

## USING THE SPINNING TOP PRINCIPLE

Expression (3) determines the probability of finding the quantum system under consideration in the initial excited state after a certain short time interval $\Delta t$. It is clear that the law of decay is not exponential. If we assume that at the time $\Delta t \ll \Delta E \cdot \hbar$ the system has been observed (the system was forced into a short-lived new state ($\tau_L < \Delta t$) and returned to its original state), then expression (3) determines the probability of finding the system in the initial state. Further, if after the time $2 \cdot \Delta t$ the system is observed again, the probability of finding it in the initial state is determined by the same formula. The decay processes at each of the intervals $\Delta t$ are independent. Therefore, the total probability of finding the system in the initial state will be determined by the product of these two probabilities. Continuing these arguments, one can obtain the following expression for the probability of finding a system in the initial excited state after the time $T = N \cdot \Delta t$ ($N \gg 1$):

$$L_N = \prod_{i=1}^{N} w_i \sim exp(\Delta t / 2T_L), \quad \lim_{\Delta t/T_L \to 0} L_N(t) \to 1, \tag{4}$$

where $w_i = \left[ 1 - (\Delta t_i / T_L)^2 \right]$; $T_L = \hbar / \sqrt{\langle (\Delta E)^2 \rangle}$.

This result corresponds to the quantum Zeno effect. Thus, [1] rigorously derives one of the two key results that form the basis of the quantum Zeno effect. As for the second result - the need to observe a quantum system (the collapse of the wave function), this procedure can be replaced by a procedure that does not require going beyond the framework of traditional quantum mechanics. Indeed, if it is necessary to keep the excited state, not to let the system to decay, then one of the possibilities to achieve this result (analogous to the process of observing the system) is to create conditions when the function $w$ undergoes some changes in time ($w = w(E,t)$). As an example, we can imagine that under the influence of an external perturbation this function periodically changes. In the simplest case, it can look like this:

$$w(E,t) = \begin{cases} w_1(E), & t \in 2n \cdot \Delta t \\ w_2(E), & t \in (2n+1) \cdot \Delta t \end{cases}. \tag{5}$$

We assume that in each time interval each of the functions $w_i(E)$ is absolutely integrable functions, and the values of the intervals themselves are small. Then the probability of finding the system in the initial state at each of these intervals is determined by the formula (3). Obviously, these probabilities are independent and, as a result, the finite probability is described by the formula (4). Thus, we obtain a result completely analogous to the result of the quantum Zeno effect.

It is useful to compare these results with those obtained within the modern interpretations of the quantum Zeno effect (see, for example, [2]). Let us describe the initial state of the quantum system by function $|\psi_0\rangle = |\psi(t=0)\rangle$, and its state at the time $t$ - $|\psi(t)\rangle$. Using the evolution operator, the expression for $|\psi(t)\rangle$ can be rewritten in the form: $|\psi(t)\rangle = exp(-i\hat{H}t/\hbar)|\psi_0\rangle$. Then the amplitude of probability of finding the system in its initial state is expressed through the scalar product of these states: $A(t) = \langle \psi_0 | \psi(t) \rangle$. Taking into account the expression for $|\psi(t)\rangle$, this amplitude can be rewritten as:

$$A(t) = \langle \psi_0 | exp(-i\hat{H}t/\hbar) | \psi_0 \rangle$$

The corresponding probability to find the system under investigation in the initial state is:

$$L(t) = |A(t)|^2 = \left| \langle \psi_0 | exp(-i\hat{H}t/\hbar) | \psi_0 \rangle \right|^2. \tag{1a}$$

Expression (1a) is equivalent to the expression (1).

At small time intervals ($Ht/\hbar \ll 1$), it is convenient to expand the exponent in a Taylor series:

$$|\psi(t)\rangle = exp(-i\hat{H}t/\hbar)|\psi_0\rangle = |\psi_0\rangle - i\hat{H}t/\hbar \cdot |\psi_0\rangle - \left(\hat{H}t/\hbar\right)^2 \cdot |\psi_0\rangle/2 + \ldots$$

Leaving only the first three terms of the expression above and taking into account $\langle \psi_0 | \psi_0 \rangle = 1$ the amplitude of the probability and the probability itself can be written as follows:

$$A(t) = 1 - i\frac{t}{\hbar}\langle \psi_0 | \hat{H} | \psi_0 \rangle - \frac{t^2}{2\hbar^2}\langle \psi_0 | \hat{H}^2 | \psi_0 \rangle$$

$$L(t) = |A(t)|^2 = \left\{ 1 - \left(\frac{t}{T_z}\right)^2 \right\}. \tag{3a}$$

Here $T_z = \hbar/\Delta$, the magnitude which is called the "Zeno time";

$$\Delta = \left[ \langle \psi_0 | \hat{H}^2 | \psi_0 \rangle - \langle \psi_0 | \hat{H} | \psi_0 \rangle^2 \right]^{1/2}.$$

Formulas (1), (1a), (3) and (3a) are equivalent. Despite their equivalence, it is easy to see that the formulas (1) and (3) contain more transparent physical parameters than formulas (1a) and (3a). Of course, these parameters also exist in formulas (1a) and (3a); however, in order to see this one needs sufficient experience in the corresponding calculations.

The results obtained above indicate the existence of common features of all decay processes. However, they refer to the transitions caused by zero oscillations (spontaneous transitions). To control these processes, it is important to know the features of the decay process, which are induced by an external perturbation. Below we investigate the properties of induced transitions. Main attention is paid to those properties that allow controlling decay processes. First of all, let us consider a two-level system. The zero level corresponds to a stationary, unexcited state. The first level corresponds to the excited state. Let now use a resonance perturbation which causes the quantum system transition from the zero level to the first one and back. As we known the perturbation theory describes such process by the following simple system of differential equations:

$$i \cdot \hbar \cdot \dot{A}_0 = V_{01} A_1; \quad i \cdot \hbar \cdot \dot{A}_1 = V_{10} A_0, \tag{6}$$

where $A_i$ - are the complex amplitudes of the wave functions.

The matrix elements of the interaction $V_{01}$ and $V_{10}$, in general, depend both on the structure of the quantum system under consideration and on the perturbation characteristics. In the simplest cases, they can be considered as equal, constant and real ($V_{10} = V_{01} \equiv V$). Suppose that the quantum system at the initial time is located in excited state. Then the solutions of equations (6) are the functions:

$$A_1 = \cos(\Omega \cdot t), \quad A_0 = \sin(\Omega \cdot t), \tag{7}$$

where $\Omega = V / \hbar$ - Rabi frequency.

It is convenient to break down the entire time interval $T = 2\pi / \Omega$ into small time intervals $\Delta t = T / n$.

Suppose, at some point in time $\Delta t$, we can estimate the position of the system under study. The probability of the system to remain in the excited state is:

$$w(\Delta t) = 1 - (\Omega \cdot \Delta t)^2. \tag{8}$$

This expression practically coincides with expression (3). After the next time interval $\Delta t$, you can again analyze the system. The probability to find it in the initially excited state is determined by the formula:

$$w(2 \cdot \Delta t) = \left(1 - (\Omega \cdot \Delta t)^2\right)^2. \tag{9}$$

Such formula reflects the fact that quantum transitions are independent in each of the time intervals $\Delta t$. Ultimately, after a large number of measurements, the probability of finding the system in the excited state can be expressed by the formula:

$$w(n \cdot \Delta t) = \left(1 - (\Omega \cdot \Delta t)^2\right)^n, \quad \lim_{n \to \infty} w(n \cdot \Delta t) = 1. \tag{10}$$

Thus, the process of observing the excited system does not allow this system to pass from the initial excited state to some other state. This fact, like the expression (4), is the content of the Zenon quantum effect. In this case the effect is derived for induced processes.

It should be noted that in the presence of an external perturbation the ground state (the state of the system at the zero level) is also unstable (excited). It is easy to show that this state can also be preserved by observing the system.

The stabilizing process itself is not discussed above. Let's look at it in the details. We do not consider a measurement process but introduce an additional perturbation. Below we show that for certain characteristics of this additional perturbation (it can be called a stabilizing perturbation) it can play the role of a measurement process. To determine the characteristics of this stabilizing perturbation, we consider a multilevel quantum system, which is described by the Hamiltonian:

$$\hat{H} = \hat{H}_0 + \hat{H}_1(t) . \tag{11}$$

The second term on the right-hand side describes the perturbation. The wave function of the system (11) obeys the Schrodinger equation, which solution will be expanded into the series in eigenfunctions of the unperturbed system:

$$\psi(t) = \sum_n A_n(t) \cdot \varphi_n \cdot \exp(i\omega_n t), \tag{12}$$

where $\omega_n = E_n / \hbar$; $i\hbar \frac{\partial \varphi_n}{\partial t} = \hat{H}_0 \varphi_n = E_n \cdot \varphi_n$.

Let's substitute (12) into the Schrodinger equation and in the usual way obtain a system of coupled equations for finding complex amplitudes $A_n$:

$$i\hbar \cdot \dot{A}_n = \sum_m U_{nm}(t) \cdot A_m, \tag{13}$$

where $U_{nm} = \int \varphi_m^* \cdot \hat{H}_1(t) \cdot \varphi_n \cdot \exp[i \cdot t \cdot (E_n - E_m)/\hbar] \cdot dq$.

Let's consider the simplest case of a biharmonic perturbation $\hat{H}_1(t) = \hat{U}_0 \cdot \exp(i\omega_0 t) + \hat{U}_1 \cdot \exp(i\omega_1 t)$. Then the matrix elements of the interaction are equal:

$$U_{nm} = V_{nm} \exp\{i \cdot t \cdot [(E_n - E_m)/\hbar + \Omega]\}, \quad V^{(k)}_{nm} = \int \varphi_n^* \cdot \hat{U}_k \cdot \varphi_m dq, \quad \Omega = \{\omega_0, \omega_1\} . \tag{14}$$

Consider the dynamics of a three-level system ($|0\rangle, |1\rangle, |2\rangle$). We assume that the frequency of the external perturbation and the eigenvalues of the energies of these levels satisfy the relations:

$$m = 1, n = 0, \quad \hbar\omega_0 = E_1 - E_0; \quad m = 2, n = 0 ; \quad \hbar(\omega_0 + \omega_1) = E_2 - E_0, \quad \hbar\omega_1 = E_2 - E_1 . \tag{15}$$

Relations (15) points to the fact that the frequency of the external perturbation $\omega_0$ is in resonance with the transitions between the zero and first levels, and the frequency $\omega_1$ is in resonance with the transitions between the first and the second levels. Using these relations in the system (13), we can reduce the system to three equations:

$$i\dot{A}_0 = A_1, \quad i\dot{A}_1 = A_0 + \mu A_2, \quad i\dot{A}_2 = \mu A_1 . \tag{1b}$$

The system (1b) can be rewritten in a somewhat different form:

$$\ddot{A}_1 + \Omega^2 A_1 = 0, \quad i\dot{A}_0 = A_1, \quad i\dot{A}_2 = \mu A_1 \quad, \tag{16a}$$

where $\Omega^2 = (1 + \mu^2)$, $\mu \equiv V_{12}/V_{10}$.

The energy levels diagram for this system is shown on Fig. 1a.

In (1b), for simplicity and convenience, we set $V_{12} = V_{21}$; $V_{10} = V_{01}$; $\dot{A}_i = dA_i/d\tau$, $\tau = V_{10} \cdot t / \hbar$. In addition, we introduced the parameter $\mu$.

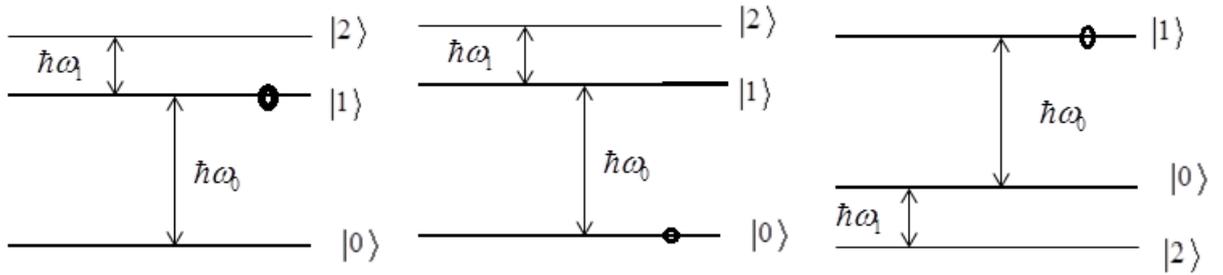

Fig.1a. Scheme of energy levels. Initially system is on excited level $|1\rangle$

Fig.1b. Scheme of energy levels. At $\tau = 0$ system is on the unexcited zero level $|0\rangle$

Fig.1c. Scheme of energy levels. At $\tau = 0$ system is on the excited level $|1\rangle$

Let's assume that at the initial moment of time ($t = 0$) the considered quantum system is at the first excited level (see Fig.1a). Then, as is easy to see, the solutions of system (1b) are the functions:

$$A_0 = \frac{1}{i \cdot \Omega} \sin(\Omega \cdot t), \quad A_1 = \cos(\Omega \cdot t), \quad A_2 = -i \sin(\Omega \cdot t). \tag{17}$$

The solution (17) shows that the larger parameter $\mu$ the smaller probability for the system to go from excited into the unexcited stationary state. It's necessary to clarify the meaning the parameter $\mu$. Physically, this parameter determines the ratio of the number of low-frequency perturbation quanta responsible for the transitions between the first and the second level to the number of the high-frequency perturbation quanta determining the transitions between the first and zero levels (Fig.1a). The greater this ratio, the lower is the probability of the excited system passing into the unexcited state.

Note that the system of equations (1b) is of the third order, and the system (16a) is of the fourth. As a result, if we solve the problem with the initial conditions $|A_0(0)|^2 = 1, |A_1(0)| = |A_2(0)| = 0$ i.e. at the initial moment of time the system is at the ground (unexcited) level (see Fig.1b) and the levels where system should transit under the influence of an external RF disturbance are involved in the fast dynamics. In this case, we write the solution of system (1b) in the form:

$$A_1 = a \cdot \exp(i \cdot \Omega \cdot t) + b \cdot \exp(-i \cdot \Omega \cdot t) .$$

Substitute the initial conditions for $A_1$ ($A_1(0) = 0$). Then the solutions of the system of equations (1b) are the functions:

$$A_1 = a \cdot [\exp(i \cdot \Omega \cdot t) - \exp(-i \cdot \Omega \cdot t)], \quad A_0 = -\frac{a}{\Omega} \cdot [\exp(i \cdot \Omega \cdot t) + \exp(-i \cdot \Omega \cdot t)] + C_0,$$

$$A_2 = -\frac{a \cdot \mu}{\Omega} \cdot \left[\exp(i \cdot \Omega \cdot t) + \exp(-i \cdot \Omega \cdot t)\right] + C_2.$$

We have three unknown constants $a, C_0, C_2$ and only two unused initial conditions. This situation occurs because the second equation of the first order in the system (1b) is replaced by the first equation of the system (16a), which is a second-order equation. Therefore, in addition to the initial conditions the solution should satisfy the second equation of the system (1b). From the initial conditions we obtain: $C_0 = 1 + a/\Omega \qquad C_2 = (a \cdot \mu)/\Omega$.

We require that these solutions satisfy equation $i\dot{A}_1 = A_0 + \mu A_2$. From this we find the following relationship between the constants $C_0$ and $C_2$: $C_0 + \mu \cdot C_2 = 0$ or: $1 + a/\Omega + (a \cdot \mu)/\Omega$. As a result, we find the value of the constant $a$: $a = -1/\Omega$. Finally, the expressions for the amplitudes of the wave functions become:

$$A_1 = -(2i/\Omega)\sin(\Omega t), \quad A_0 = 1 - \frac{1}{\Omega^2}[1 - \cos(\Omega t)], \quad A_2 = \frac{2\mu}{\Omega^2}[1 - \cos(\Omega t)] \ . \tag{17a}$$

An important and somewhat unexpected result follows from the form of the solution (17a). It consists in the fact that if the parameter $\mu$ is large, then, despite the fact that the external stabilizing effect does not affects the ground state of the system, however, this state is stabilized. As the consequence, it is possible to stabilize unstable states of quantum systems not only by acting on those states where the quantum system is located, but acting only to the states (making them dynamic) into which the system intends to transit.

Let us give one more important case, which confirms this possibility of stabilization. This is the case when the quantum system is at an excited energy level $|1\rangle$ and, under the influence of the perturbation $\hbar\omega_0 = E_1 - E_0$, it must go to a stationary level $|0\rangle$ (see Fig.1c). In addition to the disturbance at the frequency $\omega_0$, there is a stabilizing disturbance at the frequency $\hbar\omega_1 = E_0 - E_2$. The system of equations that describes such a quantum system has the form:

Приведем еще один важный случай, который подтверждает такую возможность стабилизации. Это случай, когда квантовая система находится на возбужденном энергетическом уровне $|1\rangle$ и под действием возмущения $\hbar\omega_0 = E_1 - E_0$ должна перейти на стационарный уровень $|0\rangle$ (see Fig.1c). Кроме возмущения на частоте имеется стабилизирующее возмущение на частоте $\hbar\omega_1 = E_0 - E_2$. Система уравнений, которая описывает такую квантовую систему, имеет вид:

$$i\dot{A}_1 = A_0, \quad i\dot{A}_0 = A_1 + \mu A_2, \quad i\dot{A}_2 = \mu A_0 \ .$$

$$\text{or} \qquad \ddot{A}_0 + \Omega^2 A_0 = 0, \quad i\dot{A}_1 = A_0, \quad i\dot{A}_2 = \mu A_0 \tag{6b}$$

Решением системы (6b) будут функции:

$$A_0 = \frac{1}{\Omega}\sin(\Omega \cdot t), \quad A_1 = 1 - \frac{1}{i \cdot \Omega^2}\cos(\Omega \cdot t), \quad A_2 = \frac{\mu}{i \cdot \Omega^2}[1 - \cos(\Omega \cdot t)] \tag{17b}$$

It can be seen from the form of this solution that as soon as the parameter $\mu \equiv V_{02}/V_{10} \gg 1$ becomes sufficiently large, the probability of a quantum system from an excited state $|1\rangle$ to go over to the ground stationary state $|0\rangle$ tends to zero.

## COMPARISON OF THE MECHANISM OF THE STABILIZATION OF UNSTABLE SYSTEMS IN A QUICK-OSCILLATING FIELD WITH THE SPINNING TOP PRINCIPLE

### General Considerations

Let's firstly briefly outline some general considerations that allow us to understand the stabilization mechanism, which we call the spinning top principle. In the overwhelming majority of cases, the unstable states of dynamic systems in a phase space are locally characterized by singular points of the "saddle" type. Unstable nodes and focuses are much less common. Using an example of an unstable point of the "saddle" type let us explain how such singular point can be transformed into an elliptic point (to a point of the "center" type). The phase portraits in the vicinity of the saddle point are shown in Figures 2-3. Equations on the phase plane that describe the dynamics of phase trajectories in the vicinity of a saddle point have the form:

$$\dot{x}_0 = \gamma \cdot x_1 \quad \dot{x}_1 = \gamma x_0. \tag{18}$$

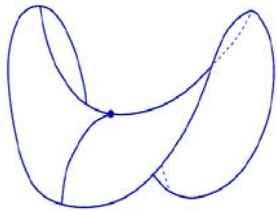
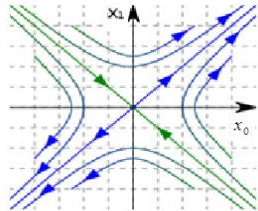
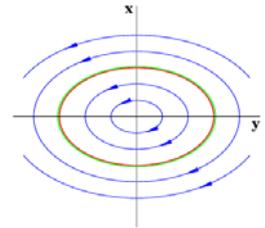

*Fig.2. Phase portrait in vicinity of the saddle point*

*Fig.3. The phase portrait in the vicinity of a singular point of the "saddle" type*

*Рис.4. The phase portrait in the vicinity of a singular point of the "center" type*

In majority of real cases, each of the dependent variables of equations (18) is some kind of characteristic of the inherent degree of freedom of the dynamical system under study. For example, these can be complex amplitudes of nonlinearly interacting waves. Therefore, in this paper, we will assume that each such first-order equation describes one of the degrees of freedom of the systems under investigation. Suppose now we need to transform the neighborhood of a saddle singular point into a neighborhood of a stable singular point, for example, an elliptic point (Fig. 4). To achieve this one can do the following operation. One introduces into the system an additional degree of freedom which is connected to one of the degrees of freedom of the unstable system. The simplest model that describes the dynamics of the system in the vicinity of the saddle point with such a modification differs from equation (18) only by one additional equation:

$$\dot{x}_0 = \gamma \cdot x_1 + \delta \cdot x_2; \quad \dot{x}_1 = \gamma x_0; \quad \dot{x}_2 = -\delta \cdot x_0. \tag{19}$$

Here, in contrast to (18), we introduce an additional degree of freedom. Moreover, this new degree of freedom is connected to one of the degrees of freedom of the original system by a coefficient $\delta$. The system of equation (19) is equivalent to the equation of a linear pendulum:

$$\ddot{x}_0 + \left(\delta^2 - \gamma^2\right)x_0 = 0 \ . \tag{20}$$

It is immediately seen from equation (20) that as soon as the coefficient describing the connection between the degrees of freedom is greater than the instability increment ($\delta > \gamma$), the unstable saddle point becomes an elliptic point. The phase space shown on Fig. 3, becomes the phase space shown on Fig. 4.

This simple way of transforming an unstable saddle point to an elliptic point well illustrates the spinning top type stabilization. Indeed, if we would not introduce an additional degree of freedom our system is unstable (the top falls). The time of the instability development ($T \sim 1/\gamma$) can be associated with the time of a top fall. The additional degree of freedom which stabilizes the system is similar to the presence of a top rotation. Moreover, spinning top principle give to us not only qualitative presentation about stabilization but it give to us also main quantitative characteristic for realization of such stabilization. Indeed, for the vertical position of a spinning top to be stable is necessary for the rotation period to be smaller than the time of a fall. In the models above (see equation (20)), in order for the system to become stable is also necessary to have coupling coefficient to be a large than the instability increment $\delta > \gamma$. If the instability increment is zero the system of equations (19) or (20) describes simple oscillations with frequency $\delta = 2\pi / T_{rot}$. As the result, there is qualitative and quantitative analogy between a spinning top stability and the stabilization mechanism described here.

Let's make the following remark. We are used to the fact that an increase in the number of degrees of freedom of a dynamical system leads to stricter conditions for achieving it steady state. Indeed if our physical system is described by the following system of equations:

$$\dot{Z}_n = F_n(\vec{Z}, t) \ . \tag{21}$$

the type of the stability of this system at a selected point in phase space is described by a linear system of equations that describes the dynamics of small deviations $\vec{x} = \vec{Z} - \vec{Z}_0$:

$$\dot{\vec{x}} = \hat{A}\vec{x} \ . \tag{22}$$

In many cases, the coefficients of the matrix can be considered as constants. Then, in order to determine the nature of a singular point at which the system (22) is written, we must find the roots of the characteristic equation:

$$\det(\hat{A} - \lambda \cdot \hat{I}) = 0; \quad \alpha_0 \lambda^n + \alpha_1 \lambda^{n-1} + \alpha_2 \lambda^{n-2} + ... + \alpha_{n-1} \lambda + \alpha_n = 0 \ . \tag{23}$$

The Routh-Hurwitz and other similar stability criteria state that the higher the order of the equations (22) and (23), the more difficult is to satisfy the conditions to achieve the stable dynamics of this system. In the example above, we increased the number of degrees of freedom. But the opposite result has been achieved. It may look like an exceptional case, however it is not. Below, and in more detail in [3-8] is shown that the introduction of an additional degree of freedom in much more complex systems such as those that describe the stabilization of radiation fluxes in a plasma the introduction of such additional degree of freedom could also lead to the stabilization of unstable states. This result can be understood if we take into account that the additional (stabilizing) degree of freedom creates a new rapid dynamics. A hierarchy and competition of processes are created [8-9]. In this case, one of the degrees of freedom of the original (unperturbed) system "drops out" of slow dynamics.

**Comparison between spinning top and the Kapitsa pendulum stabilization**

The above-considered quantum Zeno stabilization and the spinning top stabilization are based on the same rapid change in characteristics of the systems. This change resembles the change common for the dynamics of particles in a rapidly oscillating field. The simplest way of describing the motion of systems in such fields was proposed by Kapitza for analysis of the dynamics of a mathematical pendulum with rapidly oscillating point of suspension [10, 11]. His study contains the most important features of the dynamics of systems in a rapidly oscillating field.

Let's use this simple model to compare the mechanism of stabilization of the dynamics of systems in a fast-oscillating field and the spinning top stabilization to identify similarities and differences between these two systems. First of all, let us briefly describe the dynamics of a mathematical pendulum with rapidly oscillating parameters. In describing such a dynamics we use the algorithm described in Landau [12]. The equation for a mathematical pendulum has the form:

$$\ddot{x} + \left(\Omega^2 + \varepsilon \cos(\omega \cdot t)\right)\sin x = 0 . \tag{24}$$

where $\Omega$ - the eigenfrequency of small oscillations of the pendulum and $\omega$ - frequency of rapid oscillations of the pendulum parameters. In particular, this frequency can be the frequency of the change in the position of the suspension point of a mathematical pendulum. It is assumed that this frequency is much larger than the eigenfrequency of the pendulum ($\omega \gg \Omega = 2\pi/T$). It is convenient to switch to a new independent variable $\tau = \Omega t$. Then equation (24) can be rewritten:

$$\ddot{x} + \left(1 + q\cos(\omega_N \cdot \tau)\right)\sin x = 0, \tag{25}$$

where - $q = \varepsilon/\Omega^2$, $\omega_N = \omega/\Omega \gg 1$.

Following [12], we write equation (25) in the form:

$$\ddot{x} = -\frac{dU}{dx} + f(x,t), \tag{26}$$

where $\frac{dU}{dx} = \sin x$   $f(x,t) = -q \cdot \cos \omega_N \tau \cdot \sin x$.

Next, we represent the dependent variable as a sum of slowly varying ($X(t)$) and rapidly varying ($\xi(t)$) quantities: $x(t) = X(t) + \xi(t)$. After substituting this expression in equation (26) we assume that the rapidly varying quantity is small in comparison with the slowly varying quantity. We expand the functions entering the right-hand side of equation (26) in a Taylor series in a neighborhood of the function $X(t)$. Constraining ourselves by the first nonvanishing terms of this expansion, equation (26) can be rewritten as:

$$\ddot{X}(t) + \ddot{\xi}(t) = -\left[\Omega^2 \sin X + \varepsilon \xi \cos X \cdot \cos \omega t\right] - \varepsilon \sin X \cdot \cos \omega t - \Omega^2 \cos X \cdot \xi. \tag{27}$$

Applying to the left and right sides of equation (27) the averaging procedure with respect to a rapidly varying quantity, i.e. integrating over the period $\tau = 2\pi/\omega$: $\langle Z \rangle = \frac{1}{\tau}\int_0^\tau Z \cdot dt$ and noticing that the slowly varying quantities "do not notice" such averaging, we find the following expression for the rapidly varying quantity:

$$\xi = -\left(q/\omega_N^2\right)\sin X \cdot \cos \omega_N \tau \qquad (28)$$

and the equation that describes the slow dynamics of the pendulum:

$$\ddot{X} = -\frac{dU_{eff}}{dX}, \qquad (29)$$

where - $U_{eff} = -\cos X + \alpha \sin^2 X$, $\alpha = q^2/4\omega_N^2$.

The stable position of the mathematical pendulum corresponds to the minimum of the effective potential, i.e. is determined by the vanishing of the derivative of the potential: $\partial U/\partial x = \sin x \cdot [1 + 2\alpha \cdot \cos x] = 0$. One can see that the lower position of the pendulum ($x = 0$) is always stable. All those positions of the pendulum where $[1 + 2\alpha \cdot \cos x] = 0$ are also stable. In particular, the upper (vertical) position of the pendulum ($x = \pi$) is stable when the condition $q^2 > 2\omega_N^2 \gg 1$ is satisfied. In Figures 6-7 show the potential $U(x) = -\cos(x) + \alpha \cdot \sin^2(x)$ for the values $\alpha = 0.5$ (Fig. 6) and for $\alpha = 1.2$ (Fig. 7). One can see from these figures that for the small values of the parameter $\alpha$ the effective potential contains only one minimum point ($x = 0$), which corresponds to the lower position of the pendulum. However, starting from the value $\alpha > 1$, a local minimum appears at the point $x = \pi$ too. The depth of this local minimum increases with growth of the parameter $\alpha$. Both the degree of the stability and the region of stable values of the angular variable $x$ are also increasing.

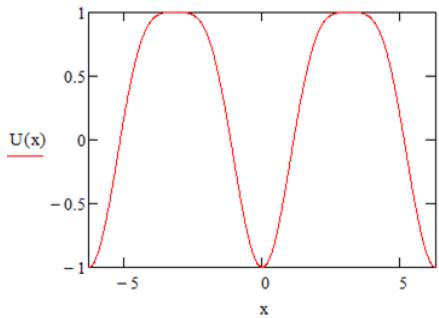
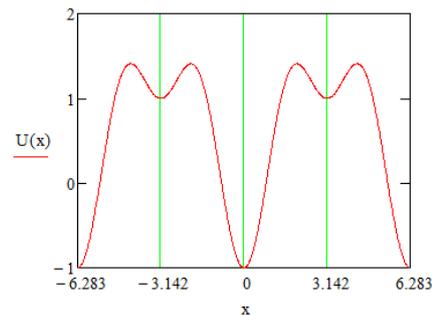

*Fig. 6. Effective potential at $\alpha = 0,5$*   *Fig. 7. Form of the effective potential at $\alpha = 1,2$*

To compare the dynamics of the Kapitza pendulum with the dynamics of the system, which is stabilized using the spinning top principle, it is convenient to directly investigate the dynamics of a mathematical pendulum in the vicinity of an unstable point $x = \pi$. The dynamics of the pendulum in the vicinity of this point can be described by a system of equations:

$$\ddot{x}_0 - \Omega^2 x_0 = \varepsilon \cos(\omega \cdot t) x_0 = x_1 \cdot x_0 \qquad (30)$$

$$\ddot{x}_1 + \omega^2 x_1 = 0, \quad x_1(0) = \varepsilon; \ \dot{x}_1(0) = 0.$$

Here $x_0 = x - \pi$

The second equation, the equation for $x_1$, describes the dynamics of the external stabilizing force. Let's pay attention to the fact that the dynamics of this force does not depend on the dynamics of the system being stabilized (from $x_0$). Therefore, in this case, the external impact is not self-consistent with the dynamics of the pendulum. This feature is typical for the stabilization of other systems when using (for stabilization) external rapidly oscillating forces. In many cases this feature is useful. Let's rewrite equation (30) through the same variables as equation (25):

$$\ddot{x}_0 - x_0 = q \cdot \cos(\omega_N \cdot \tau) x_0 = x_0 x_1 \equiv f(x_0, \tau) \qquad (31)$$

$$\ddot{x}_1 + \omega^2 x_1 = 0, \quad x_1(0) = q; \ \dot{x}_1(0) = 0.$$

The system of equations (31) considered as a system that describes the motion in a constant in time potential ($U(x) = -x^2/2$) and under the influence of an external rapidly oscillating force, which changes the parameters of the pendulum ($q \cos(\omega \cdot t) x_0 \equiv f(x_0, t)$).

$$\ddot{X} = -\frac{dU_{eff}}{dX}. \qquad (32)$$

Here $U_{eff}(x) = -\frac{x^2}{2} + \frac{1}{2\omega_N^2} \langle f^2 \rangle = -\frac{x^2}{2} \left[ 1 - \frac{q^2}{2\omega_N^2} \right]$.

The stable state corresponds to a minimum of this potential:

$$q^2 > 2\omega_N^2 \gg 1. \qquad (33)$$

Naturally, condition (33) coincides with the stability condition for the vertical position of the pendulum obtained above. As a concrete example, consider a pendulum with length $l$ oscillating in a gravitational field, and which suspension point oscillates with frequency $\omega \gg \Omega = \sqrt{l/g}$. If the maximum deviation of the suspension point is equal to $a$, the stability condition of the vertical position of the pendulum is:

$$\omega > \sqrt{2} \cdot \Omega \cdot \frac{l}{a}; \quad \Omega = \sqrt{l/g}. \qquad (34)$$

By definition (at the formulation of the problem) this is a large quantity. If we look only at the left-hand side of equation (31), this equation describes an unstable stationary point of a "saddle" type. The presence of a rapidly oscillating parametric force (the right-hand side of Eq. (31)), as we have seen, under condition (33) leads to the transformation of the saddle point to a stable point of a "center" type. Such description of stabilization of a vertical position of a mathematical pendulum is convenient for comparing it with a spinning top stabilization. Indeed, consider an unstable saddle point corresponding to the equation (31) in the absence of a right stabilizing force:

$$\dot{x}_0 = \Omega \cdot x_1 \quad \dot{x}_1 = \Omega \cdot x_0 \ . \tag{35}$$

Let's now suppose that the variable $x_0$ is connected by a linear bond with some other variable $x_2$. The system of equations that describes such a modified system can have the form:

$$\dot{x}_0 = \Omega \cdot x_1 + \delta \cdot x_2$$
$$\dot{x}_1 = \Omega \cdot x_0, \quad \dot{x}_2 = -\delta \cdot x_0. \tag{36}$$

The system of equations (36) is equivalent to the equation of a linear pendulum:

$$\ddot{x}_0 + \left(\delta^2 - \Omega^2\right)x_0 = 0. \tag{37}$$

It follows from this equation that as soon as the condition $\delta > \Omega$ is satisfied, the unstable saddle point becomes a stable point of a "center" type. Comparing this stabilization condition with the stabilization condition of the vertical position of the pendulum (33), we see that the spinning top stabilization principle to convert an unstable saddle point to a stable point is much more effective. It should be kept in mind, of course, that if we are talking about an ordinary mathematical pendulum that oscillates in a gravitational field, the use of the spinning top mechanism to stabilize the vertical position does not look physically very convenient. However, the model of a mathematical pendulum is one of the most common and most convenient models to reduce descriptions of dynamics of a large number of diverse physical systems. Therefore, in all cases where the spinning top type stabilization can be used, it turns out to be more effective than a simple parametric influence on the parameters of an unstable system. Let's note one more difference between the Kapitza's mechanism of stabilization and the use of the spinning top mechanism. This difference is easily to see if one compares the system of equations (31) and (36). In the first case the external stabilizing disturbance is an independent parameter. The oscillations of the stabilized pendulum do not affect the dynamics of the external force. In contrast, the stabilizing effect in system (36) is self-consistent. The oscillations of the stabilized system significantly affect the dynamics of the stabilizing force.

### EXAMPLES OF USING OF THE SPINNING TOP MECHANIZM FOR STABILIZATION OF UNSTABLE SYSTEMS

In this section we briefly describe some examples of the use of the spinning top stabilization mechanism. Expanded investigation of the provided examples can be found in [3-8]. In [3,7] (see also [5,8]) it was shown that the spinning top principle can be used to suppress synchrotron radiation (SR), i.e. for the stabilization of high Landau levels. Below we describe the conditions necessary to suppress the SR. Of particular interest here is the comparison between the classical suppression effect and the quantum suppression effect.

**Classical assessment of the conditions of the SR suppression.**

First of all, let us consider how the SR can be suppressed within the framework of classical electrodynamics. In this case, we assume that if the electron goes beyond the emission angle of the SR under the action of an external electromagnetic wave its radiation is suppressed. To estimate the necessary field strength, let us use the fact that the length of the SR formation can be estimated as $l \approx \lambda \gamma^2$ [13]. Here $\lambda$ - the length of the radiated wave, $\gamma$ - the energy of the particle.

The corresponding radiation angle should be $\theta \approx 1/\gamma$. The time for which the particle passes the path equal to the length of the formation is: $\tau \approx l/c = \lambda \gamma^2 / c$.

To allow the field of the external stabilizing wave to "knock out" the particle from the radiation angle, it is necessary that the frequency of this wave is larger then $\Omega > 2\pi/\tau = 2\pi c / \lambda \gamma^2$. On the other hand, the angle at which the particle moves can be estimated by magnitude $\theta \approx r_\perp / r_\parallel = r_\perp / l \ll 1$. The transverse deviation of a particle in the field of an external electromagnetic wave is estimated by the quantity $r_\perp \approx eE / m_\perp \Omega^2$. In order to suppress the radiation process, it is necessary for particle to escape (under the influence of a perturbation) beyond the emission cone: $r_\perp / l > 1/\gamma$. From these conditions it is possible to determine the necessary field strength which is equal $eE / m_0 c \Omega > 2\pi$.

**Quantum assessment of the SR suppression conditions.**

Let us now estimate the intensity of the electric field of the external electromagnetic wave necessary for stabilization if we take into account the quantum Zeno effect. In accordance with the spinning top stabilization mechanism, the first step in determining the conditions for suppression is the determination of the lifetime of the excited state. This time for SR in a synchrotron in the absence of a perturbation can be estimated by the formula $T_L = (\hbar \cdot R / r_0 \cdot mc^2 \cdot \gamma)$ [14]. Here $r_0$ - is the classical radius of the electron; $R$ - is the radius of the electron orbit in the synchrotron. For the case of usual parameters of a synchrotron: $R = 100 cm$, $E = mc^2 \cdot \gamma = 500 MeV$ and the lifetime is in the order of $10^{-9}$ sec. The second step is to find the Rabi frequency. The Schrodinger equation (Dirac) includes the magnitude of the potential of the external wave. We estimate this potential by magnitude $V \approx eE\lambda$. Accordingly, the Rabi frequency equal to $\Omega_R = V/\hbar = eE\lambda/\hbar$. We saw that for the suppression effect it is necessary to have this frequency larger than $2\pi/T_{LF}$ ($\Omega_R \gg 2\pi/T_{LF}$). From here one can obtain the following estimate for the magnitude of the field strength needed for suppressing the SR: $E > (10^{10} \hbar / e\lambda) 300 \sim (10^{-5}/\lambda)(V/cm)$. It can be seen from this estimate that in the quantum case the value of the field strength needed for stabilization is many orders of magnitude smaller than the field strength obtained in the framework of classical electrodynamics. This result is easily explained by the fact that in the quantum-mechanical analysis of suppression is only necessary for the Rabi frequency to be higher than the inverse lifetime of the electrons in the excited state. In the framework of classical electrodynamics, such processes are simply absent.

The described mechanism (spinning top principle) can be successfully used to stabilize classical systems. Below we investigate its applicability for suppression of plasma-beam instability and for suppression of decay instability in the propagation of radiation flow in a nonlinear media, in particular in plasma.

**Suppression of plasma-beam instability**

Let's consider a plasma cylinder ($0 < r < R_p$) placed in a metal shell of the same radius. The plasma is placed in a strong external longitudinal magnetic field. The beam with the radius of the metal cylinder propagates along the axis of the plasma cylinder. The metal shell contains elements of connection with an external electrodynamic structure (for example, holes (slits)). A spiral of radius $R_H$ can be chosen as the example of electrodynamic structure. Thus, we have three main oscillatory systems: plasma ($n_p$), beam ($n$) and external oscillatory structure ($E_2$). For effective interaction between the oscillations in plasma and the oscillations in an external electrodynamic system, their frequencies must coincide ($\omega_p = |k_\perp| c$). The longitudinal wave numbers must also coincide. The system of equations that describes the dynamics of such oscillatory system can be represented by the system of three coupled oscillators:

$$\ddot{n}_p + \omega_p^2 n_p = -\omega_p^2 n + i\mu \frac{k_z n_{0p} e}{m} E_2 \qquad (38)$$

$$\ddot{n} + \omega_b^2 n - 2ik_z \dot{n} - k_z^2 V^2 n = -\omega_b^2 n_p + i\mu \frac{k_z n_0 \cdot e}{m} E_2$$

$$\ddot{E}_2 - k_\perp^2 c^2 E_2 = \mu_1 \frac{4\pi e}{ik_z}(n_p + n).$$

Here $\mu_1 = \mu / G$, $G$ is the norm of the wave field in the external structure, $k_\perp^2 = (\lambda_n^2 / R_H - k_z^2)$, $E_2$ - is the longitudinal component of the electric field of the wave in the external electrodynamic structure, $\lambda_n$ - the roots of the Bessel functions ($J_0(\lambda_n) = 0$), $\mu$ - is the coupling coefficient of the plasma wave with the eigenwave of the external electrodynamic structure. From equations (38) we can obtain the following dispersion equation:

$$\left[1 - \frac{\omega_p^2}{\omega^2} - \frac{\omega_b^2}{(\omega^*)^2}\right] - \frac{\mu\mu_1}{(\omega^2 + k_\perp^2 c^2)} \left(\frac{\omega_p^2}{\omega^2} + \frac{\omega_b^2}{(\omega^*)^2}\right) = 0. \qquad (39)$$

Here $\omega^* = \omega - k_z V$, $V$ - velocity of the beam.

From this dispersion equation we see that in the absence of a connection between the fields in the plasma and in the external structure ($\mu_1 = 0$), the ordinary dispersion equation of the plasma-beam system is obtained. Similarly, if the beam is absent ($\omega_b^2 = 0$), then it converts into the dispersion equation, which describes the energy transfer between the plasma waves and the waves of the external structure. The frequency of such transfer is $\Omega = \sqrt{\mu\mu_1}/2$. In accordance with the spinning top mechanism, it can be calculated that such instability is suppressed if this frequency becomes higher than the increment of the beam instability. Analytical and numerical studies have shown that the plasma-beam instability does not develop as soon as the inequality $\sqrt{\mu\mu_1}/2 > (\omega_b^2 \omega_p / 2)^{1/3}$ is satisfied.

**Stabilization of radiation fluxes in plasma**

Decay instability develops when waves propagate in plasma. It can be useful but can be also undesirable or harmful particularly in the case when such instability leads to a stochastic regime [15-16]. In this case, it should be suppressed. Below we show that such instabilities can be suppressed by involving just one of the participating in the three-wave interaction waves in some additional periodic process (stabilizing process). The simplest system of equations that describes such processes can be written in the following form:

$$\frac{dA_0}{dt} = -VA_1 A_2 + \frac{\mu}{2i} A_3; \qquad \frac{dA_3}{dt} = \frac{\mu}{2i} A_0; \qquad (40)$$

$$\frac{dA_1}{dt} = VA_0 A_2^*; \qquad \frac{dA_2}{dt} = VA_1^* A_0.$$

This system of equations describes the interaction of four waves. Moreover, two of them, zero and the third in our notation are connected with each other by a linear bind characterized by a coupling coefficient $\mu$. If there are no additional waves a periodic transfer of energy from the main wave to the stabilizing (third) wave and back occurs. The frequency of such transfer is $\Omega = \mu/2$. Three waves (zero, first and second) interact through nonlinearity. If the coupling coefficient is zero ($\mu = 0$), then the system (40) describes the ordinary three-wave interaction of waves, the dynamics of which is well studied [17, 18]. The increment of the decay instability is $\delta = V|A_0(0)|$. Note that if the sign of the first term on the right-hand side of the first equation changes from negative to positive, then such system describes the explosive instability, which has also been studied in detail. The process of stabilization of the decay and explosive instabilities always occurs when the stabilizing wave is turned on and the condition $\mu/2V > |A_0(0)|$ is satisfied.

## Suppression of Local Instability

It will be seen below that the spinning top type stabilization can be useful for suppression of dynamic chaos regimes. The reason for appearance of dynamic chaos regimes is a local instability. In this case the phase space contains large number of saddle points. In particular case of homoclinic structures this number is infinitely large. In the vicinity of each such point closely located trajectories exponentially diverge from each other. Above we show that if one of the degrees of freedom participating in the nonlinear interaction is involved in some additional rapid process an unstable saddle point can be transformed into a stable point. One can expect that a similar transformation of saddle points can be realized in systems with dynamic chaos. To do this, you need to link at least one of the dependent variables with some additional dependent variable. Below we shall see that such process can indeed be realized. As an example, consider a somewhat modified Lorentzian model:

$$\dot{x} = \sigma(y - z) - \mu w; \quad \dot{y} = r \cdot x - y - x \cdot z; \quad \dot{z} = x \cdot y - b \cdot z; \qquad \dot{w} = \mu x. \qquad (41)$$

If the coupling coefficient is zero ($\mu = 0$) the system of equations (41) describes well known Lorentz model. When the parameters $\sigma = 10$, $b = 8/3$ and $r = 28$ this system is in a mode with dynamic chaos. The dynamics of such system has been studied in great detail. It can be expected that if any of the components of the Lorentz system ($x; y; z$) is connected with some fourth component ($w$), and the connection between these components is such that the period of energy exchange between these components is less than the inverse increment of the local instability, then the dynamics of the Lorentz system becomes complex but regular. Here we investigate system (41) numerically. Some results of this study are provided below. Figs 7-8 show the dynamics of the classical Lorentz system (without connection to an external additional component). It can be seen that the usual dynamics of a Lorentzian system is observed: the spectrum of its motion is wide and the correlation function rapidly decreases. Numerical calculations show that when an external dynamical variable is connected to the first component of the Lorentz system (41) the increase of the coupling coefficient to values up to five has little effect on its statistical characteristics. However, starting somewhere from five or six the dynamic becomes regular. This is illustrated below by the figures 9-10.

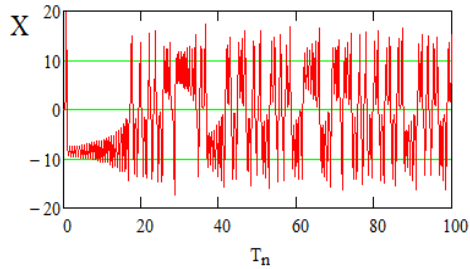

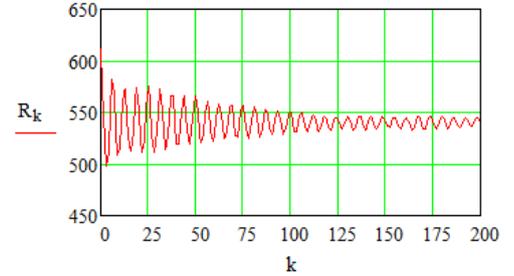

*Fig. 7. The usual dynamics of a variable Lorentz system without the influence of a stabilizing variable ( $\mu = 0$ )*

*Fig. 8. Autocorrelation function of a variable $x$ at $\mu = 0$*

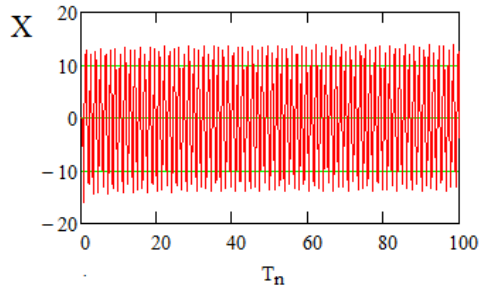

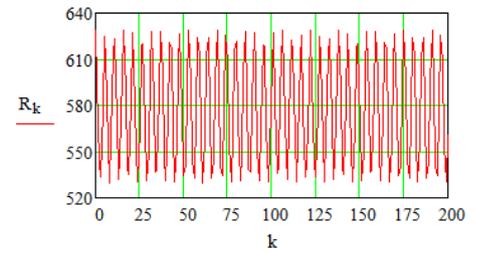

*Fig. 9. Dynamics of a variable $x$ Lorentz system in the presence of a stabilizing variable ( $\mu = 6$ )*

*Fig. 10. Autocorrelation function of a variable $x$ at $\mu = 6$*

The figures show that the dynamics became much more regular. The amplitude of the correlation function remains practically unchanged. Similar results can be obtained when an external stabilizing component is connected to two other components ( $y, z$ ) of the Lorentzian system. In all cases when the value of the coupling is sufficiently large the dynamics of the Lorentzian system becomes regular. It should be noted that a large number of various physical systems can be described by system (41). Therefore, the question arises about physical nature of the external component, which can stabilize the Lorentzian system. It is clear that in all particular cases it would be different physical variable. In particular, we point out that if the Lorentz system describes the dynamics of a single-mode laser, the dependent variable $x$ determines the amplitude of the laser field. In this case, it is easy to imagine the practical implementation of the suppression mechanism. It suffices to connect the field of a single-mode laser to a certain field of another mode. The connection can be linear or nonlinear. Chaotic dynamics becomes a regular dynamic if the connection turns out to be sufficiently large. One should note that the coupling coefficient in this case is much larger than it was in previous cases ( $\mu = 6$ ). One could ask, what is the value of the coupling factor necessary for suppressing modes with dynamic chaos? In fact, at each point of the phase space the divergence of the phase trajectories is different (the maximal Lyapunov exponent is different). The analysis of all such points, in general, is unproductive so it makes sense, apparently, to be guided by the maximum and minimum values of Lyapunov's exponents. As an example let's find the maximum Lyapunov's index for unperturbed Lorentzian system. The equation for this index has the form:

$$\lambda^3 + \lambda^2(1+b) - \lambda\left[(b+x_0^2) + \sigma(r-z_0) - y_0\right] + \sigma\left[(r-z_0)(b-x_0) - y_0 - x_0 \cdot y_0\right] = 0.$$

Here $\{x_0, y_0, z_0\}$ - are the coordinates of the point in the phase space in which the Lyapunov exponents are determined.

An analysis of this equation shows that in the neighborhood of the zero stationary point $\{0;0;0\}$ the maximal Lyapunov's exponent is around 13. In the neighborhood of stationary point $\{\sqrt{r-1}; \sqrt{r-1}; r-1\}$ it has order of 4.8, and near the stationary point $\{-\sqrt{r-1}; -\sqrt{r-1}; r-1\}$ it is close to 6.6. Maximal Lyapunov exponent for larger values of phase coordinates e.g. $\{20;20;20\}$ is near 21. Thus, we see that in the majority of the analyzed points Lyapunov's exponents are quite large. However, to suppress the dynamic chaos, it is sufficient to introduce a coupling parameter equal to 6 ($\mu = 6$).

## Results of experimental investigations

At present, considerable amount of experimental results confirms the existence of the quantum Zeno effect [19-22]. Let us briefly outline the main ideas of these experiments and some of their results. Basically, three-level systems are considered. In most cases, the distance between the lower (first) and the second energy levels is in the radio frequency range. The energy of the third level is much higher than the energy of the lower two levels. Transitions between the third and the second levels are prohibited. In the vast majority of cases, the transitions between the first and the third levels are in the optical range. Transitions between the first and the second levels are in the microwave-frequency ranges. At the initial moment of time only the lower main level is populated. Microwave radiation with the frequency corresponding to the transitions between the first and second levels is applied to the quantum system under investigation. The characteristic transition time between the first and the second energy level selected to be as long as possible. In particular in [22] a virtual level exists between the first and the second level and the transition between these levels is driven by two microwave-frequency pulses.

A short time after the exposure to the microwave-frequency signal the system is illuminated by short laser $\pi$- impulse. This impulse quickly drives the system from the first energy level to the high energy level (the third one). The lifetime at this level is very short and in most cases this can be neglected. Therefore, the system from the third level almost instantly returns back to the first initial level. It is assumed that such laser pulse impact is a process of measurement of the state of the system. If the number of such laser pulses occurring during the transition time from the first to the second level is high than the population of the first level remains practically unchanged. The quantum Zeno effect was first discovered using such scheme of experiment and this scheme differs only in insignificant details in other quantum Zeno experiments. In all publications, the discussion about the possibilities of using and implementing the quantum Zeno effect highlights the statement that such processes cannot be realized at the nuclear level, for example, to suppress or accelerate radioactive decay processes.

Two series of experiments were conducted at the NSC "KIPT"[1], the results of which can be explained on the basis of the spinning top stabilization mechanism [23-24] (see also [5,8]). The first series of experiments [23] investigated a photocathode of a photomultiplier placed in a microwave field. A reliable, well-observed decrease of the photocurrent was observed when the microwave field was applied to the photocathode. The dependence of the decrease of the photocurrent on the power of the high-frequency field is in good qualitative agreement with the theoretical predictions. The greater was the power of the high-frequency field, the smaller was the photocurrent. This experiment is interesting because the stabilizing field acts not on the state of the electrons in the potential well of the photocathode but on the states at which the electron would transit as a result of absorption of the optical quantum. It should, however, be noted that at present there is no rigorous proof that such a decrease in the photocurrent is indeed related to the spinning top-type stabilization of the photoemission. Indeed, it is difficult to estimate the entire dynamics of electrons that move in the high-frequency field and hitting the electrodes of the photomultiplier.

The second series of experiments [24] is related to nuclear processes. The essence of the experiments conducted at the NSC "KIPT" and the Kiev Institute for Nuclear Research was as follows. A sample of the material that contained the radioactive isomer of hafnium-178 was influenced by electron flow with energies from 10 keV to 50 keV. The sample was acting as a target to convert the electrons energy into the energy of X-ray radiation. There is assumption that a short-lived trigger level somewhat higher than the basic metastable state of hafnium-178 exists and the purpose of the experiments was to drive the hafnium-178 from its metastable state to an unstable short-lived trigger level. On the experiment success an ideal nuclear fuel would appear in the hands of researchers. Indeed, a distinct acceleration of the radioactive decay process was observed in many experiments and was attributed to the influence of the electron flux to the target containing the radioactive isomer. However, more often the acceleration the authors observed the opposite effect, when the process of radioactive decay was inhibited by the action of the beam on the target. The results of this series of experiments can be explained by the mechanism of stabilization of excited states formulated above. Indeed, in the nuclear system under investigation in addition to the trigger level there are large numbers of other long-lived energy states. When a target is irradiated by an electron beam a wide spectrum of X-ray radiation is excited. Such radiation drives the nuclear system not only into a certain energy (preferably trigger) level, but also to many other long-lived levels. The system of equations that describes such transitions can be written as:

$$i\dot{A}_0 = A_1, \quad i\dot{A}_1 = A_0 + \sum_{k=2}^{N} \mu_k A_k, \quad i\dot{A}_2 = \mu_2 A_1, \quad i\dot{A}_3 = \mu_3 A_1, \ldots i\dot{A}_N = \mu_N A_1. \qquad (42)$$

This system of equations describes the connection among levels and spontaneous transitions of the system under consideration from the first metastable level ($A_1$) to the main level ($A_0$) and induced transitions to other levels close to the metastable level. The system (42) has the analytical solution:

$$A_1 = \cos(\Omega \cdot t), \quad A_0 = \frac{1}{i \cdot \Omega}\sin(\Omega \cdot t), \quad A_k = \frac{\mu_k}{i \cdot \Omega}\sin(\Omega \cdot t), \quad k = 2,3,\ldots N,$$

---

[1] http://www.kipt.kharkov.ua/en

where $\Omega = \sqrt{1 + \sum_{k=2}^{N} \mu_k^2}$.

The structure of these solutions shows that the greater is the number of energy levels to which an induced transition is possible the smaller is the probability of spontaneous decay and the smaller the probability of a system transition to a trigger short-lived level.

## CONCLUSION. THE DISCUSSION OF THE RESULTS

In conclusion, we formulate and discuss the most important results that follow from the above comparative analysis of three stabilization mechanisms.

1. The results obtained above demonstrate existence of a sufficiently effective and simple mechanism that allows unstable states stabilization. This mechanism is a spinning top principle. It is simple, obvious and universal enough to be used both for stabilizing unstable quantum systems and for stabilizing unstable classical systems. Moreover, the analysis of this mechanism gives not only qualitative but in many cases also provides the necessary quantitative characteristics of stabilization. Also important is the fact that stabilization based on the use of the spinning top principle is suitable not only for systems with a slow decay law (non-exponential decay law), but also to control unstable systems for with the usual exponential decay law and even for the stabilization of explosive instabilities. Concerning the latter, it should be noted that in some cases numerical calculations show that complete stabilization does not occur. There is only a delay, sometimes a significant one in the development of the instability process. This problem is currently under active investigation.

2. As we saw above, the stabilization of quantum systems occurs primarily due to the fact that the decay of quantum systems at small time intervals is nonexponential. Apparently, this result was first obtained in [1]. However, up to the present time the author does not aware of any references to this work in any publication related to the analysis of the quantum Zeno effect. The second fundamental process that lies behind the Zeno effect is a measurement process. The description of a measurement process in quantum mechanics goes beyond traditional quantum mechanics. Until now, there is no consensus among specialists who are involved in the fundamentals of quantum mechanics about the role the existing mathematical apparatus that describes the measurement process is playing. A vivid example of such a mixed opinion is the results of the poll that was conducted among participants of the conference "Quantum Physics and the Nature of Reality", which was held in July 2011 in Austria [25]. Organizers asked the question: "How would you characterize the existing state of the theory that describes the measurement process?" The answers diverged from the full approval of the existing theory to the complete denial of its usefulness.

3. It should be noted that many authors pointed to the fact that existing experimental results, which confirm the results of the theory of the quantum Zeno effect can be explained within the framework of traditional quantum mechanics. Thus, in the work of Prigogine and co-authors [26], the authors, within the framework of traditional quantum mechanics, succeeded in formulating the properties of a projection operator, which turned out to be sufficient to explain the main experimental results obtained by that time with respect to the quantum Zeno effect. As we showed above, the stabilization of quantum systems can be fully described within the framework of the traditional quantum mechanics.

4. It should be noted that the known experimental studies of the Zeno effect stabilization were performed using laser irradiation of quantum transitions located in a microwave energy range, i.e. the stabilization of low-frequency transitions has been achieved using high-frequency radiation. An important result obtained above is the fact that it is possible to stabilize high-frequency transitions (laser, X-ray, etc.) by using low-frequency fields. The necessary condition for such stabilization is the condition that the frequency of the Rabi of the low-frequency transitions to be larger and better significantly large than the value that is equal to the inverse lifetime of the excited states. This means that the number of low-frequency quanta must be large enough.

5. In our opinion, a very important and unusual result is the result represented by the formula (17a,17b). See also Fig.1b and Fig.1c. It importance based on the fact that while for the quantum effect of Zeno one have to measure (influence) the state in which the quantum system is located in the general case stabilization can be achieved by influencing the states in which the system intends to fall. Thus, the stabilization of a quantum system can be achieved either by transforming a stationary unstable state into a stable dynamic state, or by transforming into dynamical states the states in which the system intends to fall. This result is especially important for the stabilization of the states where external forces are either difficult or simply impossible to apply. Examples are the photo-effect and the nuclear beta decay.

6. The stabilization of classical systems using the spinning-top type stabilization has also proved to be quite effective. It turned out that there is a difference between the mechanism of stabilization of unstable classical systems by influencing them or their parameters in an external high-frequency field and stabilizing on the basis of spinning-top type stabilization. The main difference was the fact that when the system is stabilized in a rapidly oscillating field, the stabilizing force is not self-consistent. The dynamics of the stabilizing force does not depend on the dynamics of the system being stabilized. When spinning top-type stabilization occurs, the stabilizing force is self-consistent. Its dynamics essentially depends on the dynamics of the system being stabilized, so their dynamics are self-consistent. As a result, in cases where the spinning top-type stabilization can be used to stabilize the system, it turns out to be more effective than the stabilization in a rapidly oscillating field. This result was demonstrated by the example of stabilization of the upper (unstable) position of the mathematical pendulum (the Kapitsa pendulum). However, it is obvious that this result is also true in other cases. It should be noted, however, that the stabilization mechanism by means of an external rapidly oscillating field is in some sense more universal. It is easier to implement and, it seems, can be used for a wider range of physical systems than the spinning-top type stabilization. Indeed, it is quite difficult to imagine a physical realization of the spinning top-type stabilization for a mathematical pendulum, while the physical model of Kapitza's pendulum works effectively.

7. Attention should also be paid to the fact that the suppression of certain instabilities can be explained not only by the spinning-top type mechanism, but also by some other mechanisms. Thus, if we consider the suppression of the beam-plasma instability by arranging an additional coupling of the excited plasma wave with the wave of the external electrodynamic structure (38) - (39), then the instability breakdown can be explained by the fact that such an interaction leads to the splitting of the dispersion branch of the plasma wave. A gap with the width proportional to the coupling coefficient between these waves appears in the region where the branches of the plasma wave and the waves of the external electrodynamic structure intersect. As the result, the beam mode does not interact with the branch of the plasma mode. This means that the conditions for the Cerenkov's mechanism of radiation of beam particles in the plasma disappear. This, in turn, leads to a violation of the conditions necessary for the development of plasma-beam instability [27]. An analogous explanation can also be given for the process of suppression of the decay instability. In this case, the coupling of one of the waves participating in the three-wave interaction violates the synchronism condition between them. This violation of synchronism leads to the breakdown of the decay instabilities. These examples can be continued. The attractiveness of the description on the basis of spinning top type stabilization lays in the fact that it unites all these at first glance very different processes and allows you to look at all of them from a single point of view.

8. Talking about the suppression of regimes with dynamic chaos, attention should be paid to the difficulties that can be encountered in determining specific physical mechanisms that would be equivalent to the mechanism of a spinning top rotation. Indeed, if it is easy to find such a mechanism for the Lorentz system in the case when it describes the dynamics of the laser field, but it is difficult to do it for a particle moving in the Henon-Heilis potential. Indeed, it is not difficult to use the spinning top stabilization principle formally, and to break the regime with dynamic chaos for particles that move in a crystal (in the Henon-Heilis potential [28, 29]). However the difficulties appear as soon as the question arises about the physical meaning of the additional stabilizing degree of freedom. Let's make one more remark. Very often, regimes with dynamic chaos are analyzed using the Poincaré section. However, this method does not always give the right result. Indeed, if we look at the Poincaré section for the unperturbed Lorentz system and for the perturbed one, we see that the phase volume of these points has decreased (in the system being stabilized). However, it is difficult to decide if the dynamics have become regular using this picture. For this reason we provided neither a kind of phase portrait nor kind of Poincaré sections as they only indicate a tendency towards regularization.

The author is grateful to Academ. D.M. Vavriv, Prof. A.L. Sanin and Academ. N.F. Shulga for discussions of the results and useful remarks.

# КВАНТОВЫЙ ЭФФЕКТ ЗЕНОНА, МАЯТНИК КАПИЦЫ И ПРИНЦИП ЮЛЫ. СРАВНИТЕЛЬНЫЙ АНАЛИЗ
## В.А. Буц

ННЦ «Харьковкий физико-технический институт НАН Украины» 61108, Харьков, Академическая, 1; Харьковский национальный университет им. В.Н. Каразина 61022, Украина, г. Харьков, пл. Свободы 4, Украина; Радиоастрономический институт НАН Украины ул. Искусств 4, г. Харьков, 61002, Украина
*e-mail: vbuts@kipt.kharkov.ua*

Дан сравнительный анализ трех механизмов стабилизации неустойчивых состояний физических систем. Этими механизмами являются: квантовый эффект Зенона, стабилизация неустойчивых состояний во внешнем быстроосциллирующим поле (на примере маятника Капицы), а также алгоритм, который был назван принципом юлы. Определены общие черты этих механизмов, а также различия между ними. В частности, показано, что стабилизация квантовых систем возможна без привлечения такого понятия как коллапс волновой функции. Для стабилизации достаточно, чтобы поток стабилизирующего излучения был таким, чтобы частота Раби переходов в этом поле, была по возможности большей, чем некоторая частота. Этой частотой является величина, которая обратно пропорциональна времени жизни стабилизируемого состояния. Показано, что принцип юлы позволяет стабилизировать неустойчивые состояния действуя не на сами эти состояния, а на те сосстояния, в которые неустойчивая система должна перейти.
          Показано, что стабилизация неустойчивых состояний путем воздействия на нее быстроосциллирующей силы происходит несамосогласованным воздействием, т.е. динамика стабилизирующего поля не зависит от динамики стабилизируемого состояния. Стабилизация при использовании принципа юлы происходит самосогласованными силами. В результате, во многих случаях, стабилизация с использованием принципа юлы может быть эффективнее.
**КЛЮЧЕВЫЕ СЛОВА**: квантовый эффект Зенона, стабилизация, маятник Капицы

Квантовый эффект Зенона и механизм стабилизации маятника с перевернутым подвесом (маятник Капицы) – два известных механизма, позволяющих, при создании условий их реализации, сделать устойчивыми первоначально неустойчивые стационарные состояния многих физических систем. Общей характерной чертой этих механизмов является необходимость в быстром, периодическом воздействии на неустойчивую систему. Для квантового эффекта Зенона это воздействие заключается в частом наблюдении за неустойчивой квантовой системой. Неустойчивость вертикального положения математического маятника может быть подавлена быстрым изменением положения подвеса маятника. Это хорошо известные примеры стабилизации неустойчивых квантовых и классических систем. Можно сформулировать некоторый другой алгоритм стабилизации неустойчивых состояний. Этот алгоритм заключается в том, что неустойчивая система вовлекается в некоторое ***дополнительное быстрое периодическое*** движение. При этом можно указать на свойства этого движения, которые необходимы, чтобы система из неустойчивой стала устойчивой. Отметим, что рассматриваемый алгоритм имеет простой и наглядный образ, который отображает наиболее важные характеристики этого алгоритма. Этим образом является детская игрушка – юла. Вертикальное положение юлы без вращения неустойчиво. Причем, время ее падения (время жизни вертикального положения юлы) можно сопоставлять со временем жизни неустойчивого состояния. Если же юлу привести во вращение и период этого вращения будет значительно меньшим чем время жизни вертикального положения, то это положение будет устойчивым. По аналогии с этим образом такой алгоритм подавления неустойчивостей был назван принципом (механизмом) юлы. Видно, что все три механизма имеют общую характерную черту – быстрое периодическое изменение характеристик стабилизируемых систем. Имеются, однако, и различия. Анализу этих различий и посвящена данная работа. Ниже, прежде всего, рассматривается квантовый эффект Зенона. Далее, на примере маятника Капицы, приведено сравнение принципа юлы с механизмом стабилизации во внешнем быстроосциллирующем поле. Определено различие этих механизмов. Оно заключается в том, что в принципе юлы стабилизирующее возмущение является самосогласованным с динамикой стабилизируемой системы. Стабилизирующее возмущение в маятнике с перевернутым подвесом является независимым. Показано, что в тех случаях, когда может быть использован принцип юлы, этот механизм оказывается более эффективным, чем механизм стабилизации внешним быстроосциллирующим полем. Краткое описание примеров использования принципа юлы для стабилизации различных физических систем содержится в четвертом разделе. В заключении обсуждаются основные результаты работы.

**СТАБИЛИЗАЦИЯ КВАНТОВЫХ СИСТЕМ. КВАНТОВЫЙ ЭФФЕКТ ЗЕНОНА**

Квантовому эффекту Зенона посвящена обширная литература. Достаточно включить интернет, чтобы найти как описание самого эффекта, так и описание новейших экспериментальных наблюдений этого эффекта. В качестве примера можно указать на сайт https://en.wikipedia.org/wiki/Quantum_Zeno_effect и на литературу, которая там приведена. Здесь только отметим, что в основе эффекта Зенона лежат два фундаментальных процесса: 1. Неэкспоненциальный закон распада квантовых возбужденных состояний на малых интервалах времени и 2. Коллапс волновой функции в процессе измерения. Ниже мы обращаем внимание на то, что первая фундаментальная особенность квантовых систем была описана еще в 1947 году В.А. Фоком и Н.С.Крыловым [1]. Второй процесс - процесс измерения (а, соответственно, коллапс волновых функций) не может быть описан в рамках традиционной квантовой механики. Однако для стабилизации неустойчивых состояний, как будет видно ниже, процесс измерения необязателен. Достаточно учесть некоторое внешнее возмущение, которое будет достаточно быстро и периодически переводить систему из одного состояния в другое и обратно. Такой процесс вполне описывается в рамках традиционной квантовой механики. Поэтому ниже процесс измерения (коллапс волновой функции) мы рассматривать не будем. Отметим, что часто процесс внешнего воздействия, который переводит систему в другое состояние отождествляют с процессом измерения.

В работе [1] авторы получили общее универсальное выражение для зависимости вероятности нахождения квантовых систем в начальных (исходных) состояниях от времени ($L(t)$). Это выражение известно как теорема Крылова – Фока и имеет вид:

$$L(t) = \left| \int w(E) \cdot \exp(-i \cdot E \cdot t / \hbar) \cdot dE \right|^2, \qquad (1)$$

где $w(E)$ - дифференциальная функция распределения начального состояния (плотность функции распределения); $w(E) \cdot dE$ - энергетический спектр начального состояния.

Авторы провели достаточно подробный анализ этого выражения. Отметим некоторые из результатов, которые следуют из анализа выражения (1). Прежде всего, под модулем в выражении (1) стоит характеристическая функция. Поэтому она ограничена. Более того, в нашем случае она меньше или равна единице ($L(t) = |p(t)|^2 \leq 1$). Если плотность вероятности является абсолютно интегрированной функцией, то легко видеть, что вероятность на бесконечности стремится к нулю ($\lim_{t \to \infty} L(t) \to 0$). Наиболее интересным является особенность поведения функции при малых временах. Функцию $L(t)$, как всякую характеристическую функцию, можно представить в виде ряда по моментам:

$$L(t) = \left| \sum_{n=0}^{\infty} \left( \frac{i \cdot t}{\hbar} \right)^n \cdot \frac{M_n}{n!} \right|^2 . \qquad (2)$$

Здесь $M_n \equiv \langle E^n \rangle = \int_{-\infty}^{\infty} E^n w(E) dE$.

Ограничиваясь первыми тремя слагаемыми в сумме при малых временах, получим:

$$L(t) = 1 - \left( \frac{t}{\hbar} \right)^2 \langle (\Delta E)^2 \rangle , \qquad (3)$$

где $\Delta E \equiv E - E_0$, $M_1 \equiv \langle E \rangle \equiv E_0 = \int_{-\infty}^{\infty} E w(E) dE$.

Это выражение совпадает с тем выражением, что получено в работе [1].

### Использование принципа юлы

Выражение (3) определяет вероятность найти рассматриваемую квантовую систему в исходном возбужденном состоянии по истечении некоторого малого времени $\Delta t$. Видно, что закон распада при этом не является экспоненциальным. Если предположить, что в момент времени $\Delta t \ll \Delta E \cdot \hbar$ над системой было проведено наблюдение (система вынуждено переведена в некоторое короткоживущее ($\tau_L < \Delta t$) новое состояние, и вернулась в исходное состояние), то выражение (3) определяет вероятность найти систему в исходном состоянии. Далее, если по истечении времени $2 \cdot \Delta t$ над системой опять провести наблюдение, то вероятность нахождения ее в исходном состоянии будет определяться той же формулой. Процессы распада на каждом из интервалов $\Delta t$ независимы. Поэтому полная вероятность найти систему в исходном состоянии будет определяться произведением этих двух вероятностей. Продолжая эти рассуждения, можно получить следующее выражение для вероятности найти систему в исходном возбужденном состоянии по истечении времени $T = N \cdot \Delta t$ ($N \gg 1$):

$$L_N = \prod_{i=1}^{N} w_i \sim exp(\Delta t / 2 T_L), \qquad \lim_{\Delta t / T_L \to 0} L_N(t) \to 1, \qquad (4)$$

где $w_i = \left[ 1 - (\Delta t_i / T_L)^2 \right]$; $T_L = \hbar / \sqrt{\langle (\Delta E)^2 \rangle}$.

Этот результат соответствует квантовому эффекту Зенона. Таким образом, в работе [1] строго был получен один из двух ключевых результатов, которые лежат в основе квантового эффекта Зенона. Что касается второго результата - необходимости проведения наблюдения над квантовой системой (коллапса волновой функции), то эта процедура может быть заменена процедурой, которая не требует выхода за рамки традиционной квантовой механики. Действительно, если необходимо сохранить возбужденное состояние, не дать системе распасться, то одной из возможностей добиться этого результата (аналог процесса наблюдения над системой) является создание условий, когда функция $w$ будет претерпевать некоторые изменения во времени ($w = w(E,t)$). В качестве примера можно представить, что под действием внешнего возмущения эта функция периодически меняется. В простейшем случае она может выглядеть следующим образом:

$$w(E,t) = \begin{cases} w_1(E), \ t \in 2n \cdot \Delta t \\ w_2(E), \ t \in (2n+1) \cdot \Delta t \end{cases}. \qquad (5)$$

Будем считать, что на каждом временном интервале каждая из функций $w_i(E)$ абсолютно интегрируемые функции, а величины самих интервалов малы. Тогда вероятность найти систему в исходном состоянии на каждом из этих интервалов будет определяться формулой (3). Очевидно, что эти вероятности независимы и, как результат, конечная вероятность будет описываться формулой (4). Таким образом, мы получим результат полностью аналогичный результату квантового эффекта Зенона.

Полезно сравнить полученные результаты с теми, которые получаются при современном изложении квантового эффекта Зенона (смотри, например, [2]). Обозначим начальное состояние квантовой системы как $|\psi_0\rangle = |\psi(t=0)\rangle$, а ее состояние в момент времени $t$ - $|\psi(t)\rangle$. Используя оператор эволюции, выражение для $|\psi(t)\rangle$ можно переписать в виде $|\psi(t)\rangle = exp(-i\hat{H}t/\hbar)|\psi_0\rangle$). Тогда амплитуда вероятности найти систему в ее начальном состоянии будет выражаться в виде скалярного произведения этих состояний: $A(t) = \langle \psi_0 | \psi(t) \rangle$. С учетом выражения для $|\psi(t)\rangle$ эту амплитуду можно переписать в виде:

$$A(t) = \langle \psi_0 | exp(-i\hat{H}t/\hbar) | \psi_0 \rangle.$$

Соответствующая вероятность обнаружить исследуемую систему в исходном состоянии будет иметь вид:

$$L(t) = |A(t)|^2 = \left| \langle \psi_0 | exp(-i\hat{H}t/\hbar) | \psi_0 \rangle \right|^2. \qquad (1а)$$

Выражение (1а) эквивалентно выражению (1).

На малых интервалах времени ($Ht/\hbar \ll 1$) экспоненту удобно разложить в ряд Тейлора:

$$|\psi(t)\rangle = exp(-i\hat{H}t/\hbar)|\psi_0\rangle = |\psi_0\rangle - i\hat{H}t/\hbar \cdot |\psi_0\rangle - \left(\hat{H}t/\hbar\right)^2 \cdot |\psi_0\rangle / 2 + ...$$

Оставим в этом ряду только первые три члена. Кроме того учтем, что $\langle \psi_0 | \psi_0 \rangle = 1$. Тогда для амплитуды вероятности и для самой вероятности можно написать такие выражения:

$$A(t) = 1 - i\frac{t}{\hbar}\langle \psi_0 | \hat{H} | \psi_0 \rangle - \frac{t^2}{2\hbar^2}\langle \psi_0 | \hat{H}^2 | \psi_0 \rangle$$

$$L(t) = |A(t)|^2 = \left\{ 1 - \left(\frac{t}{T_z}\right)^2 \right\} . \qquad (3а)$$

Здесь $T_z = \hbar / \Delta$ -величина, которая получила название «время Зенона»,
$\Delta = \left[ \langle \psi_0 | \hat{H}^2 | \psi_0 \rangle - \langle \psi_0 | \hat{H} | \psi_0 \rangle^2 \right]^{1/2}$.

Формулы (1), (1а), (3) и (3а) эквивалентны. Несмотря на их эквивалентность легко увидеть, что формулы (1) и (3) содержат более прозрачные физические параметры, чем формулы (1а) и (3а). Эти параметры, конечно, есть и в формулах (1а) и (3а) однако, чтобы это увидеть - нужен достаточный опыт в соответствующих вычислениях.

Полученные выше общие результаты указывают на существование общих закономерностей процесса распада. Однако они относятся к переходам, которые вызваны нулевыми колебаниями (спонтанные переходы). Для управления этими процессами важно знать особенности процесса распада, которые индуцированы внешним возмущением. Ниже рассмотрены особенности индуцированных переходов. Внимание будет обращено на те особенности таких переходов, которые позволяют использовать их для управления процессами распада. Прежде всего, рассмотрим двухуровневую систему. Нулевой уровень соответствует стационарному, невозбужденному состоянию. Первый уровень соответствует возбужденному состоянию. Пусть, теперь под действием резонансного возмущения рассматриваемая квантовая система переходит с нулевого уровня на первый и обратно. Как известно, в рамках теории возмущений такой процесс описывается следующей простой системой дифференциальных уравнений:

$$i \cdot \hbar \cdot \dot{A}_0 = V_{01} A_1; \quad i \cdot \hbar \cdot \dot{A}_1 = V_{10} A_0, \qquad (6)$$

где $A_i$ - комплексные амплитуды волновых функций.

Матричные элементы взаимодействия $V_{01}$ и $V_{10}$, в общем случае, зависят как от структуры рассматриваемой квантовой системы, так и от характеристик возмущения. В простейших случаях их можно считать равными, постоянными и действительными. Пусть, в начальный момент времени квантовая система находится в возбужденном состоянии. Тогда решениями уравнений (6) будут функции:

$$A_1 = \cos(\Omega \cdot t), \quad A_0 = \sin(\Omega \cdot t), \qquad (7)$$

где $\Omega = V/\hbar$ - частота Раби.

Удобно для дальнейшего весь интервал времени $T = 2\pi/\Omega$ разбить на небольшие временные интервалы $\Delta t = T/n$. Пусть, в момент времени $\Delta t$ каким-то образом можно оценить положение изучаемой системы. Вероятность того факта, что она за время $\Delta t$ не перейдет из возбужденного состояния в основное будет равна:

$$w(\Delta t) = 1 - (\Omega \cdot \Delta t)^2 \ . \qquad (8)$$

Это выражение практически совпадает с выражением (3). По истечении следующего интервала времени $\Delta t$ можно снова провести анализ системы. Вероятность обнаружения ее в первоначально возбужденном состоянии будет определяться формулой:

$$w(2 \cdot \Delta t) = \left(1 - (\Omega \cdot \Delta t)^2\right)^2. \qquad (9)$$

Такая формула отражает факт независимости квантовых переходов в каждом из временных интервалов $\Delta t$. В конечном счете, после большого числа измерений вероятность нахождения системы в возбужденном состоянии выразится формулой:

$$w(n \cdot \Delta t) = \left(1 - (\Omega \cdot \Delta t)^2\right)^n, \quad \lim_{n \to \infty} w(n \cdot \Delta t) = 1. \qquad (10)$$

Таким образом, процесс наблюдения за возбужденной системой не дает этой системе перейти из исходного возбужденного состояния в какое-либо другое состояние. Этот факт, также как и выражение (4), составляет содержание квантового эффекта Зенона. В данном случае для индуцированных процессов.

Следует заметить, что при наличии внешнего возмущения основное состояние (состояние системы на нулевом уровне) также является неустойчивым (возбужденным). Легко показать, что это состояние также может быть сохранено путем наблюдения над системой.

Выше не обсуждался сам стабилизирующий процесс. Рассмотрим его. Процесс измерения обсуждать не будем, а введем в рассмотрение кроме дестабилизирующего возмущения добавочное возмущение. Ниже покажем, что при определенных характеристиках этого добавочного возмущения (его можно назвать стабилизирующим возмущением) оно может играть роль процесса измерения. Для определения характеристик этого стабилизирующего возмущения рассмотрим многоуровневую квантовую систему, которая описывается гамильтонианом:

$$\hat{H} = \hat{H}_0 + \hat{H}_1(t) \ . \qquad (11)$$

Второе слагаемое в правой части описывает возмущение. Волновая функция системы (11) подчиняется уравнению Шредингера, решение которого будем искать в виде ряда по собственным функциям невозмущенной задачи:

$$\psi(t) = \sum_n A_n(t) \cdot \varphi_n \cdot \exp(i\omega_n t), \qquad (12)$$

где $\omega_n = E_n/\hbar$; $\quad i\hbar \dfrac{\partial \varphi_n}{\partial t} = \hat{H}_0 \varphi_n = E_n \cdot \varphi_n$.

Подставим (12) в уравнение Шредингера и обычным образом получим систему связанных уравнений для нахождения комплексных амплитуд $A_n$:

$$i\hbar \cdot \dot{A}_n = \sum_m U_{nm}(t) \cdot A_m, \qquad (13)$$

где $U_{nm} = \int \varphi_m^* \cdot \hat{H}_1(t) \cdot \varphi_n \cdot \exp[i \cdot t \cdot (E_n - E_m)/\hbar] \cdot dq$.

Рассмотрим наиболее простой случай бигармонического возмущения $\hat{H}_1(t) = \hat{U}_0 \cdot \exp(i\omega_0 t) + \hat{U}_1 \cdot \exp(i\omega_1 t)$. Тогда матричные элементы взаимодействия приобретут следующее выражение:

$$U_{nm} = V_{nm} \exp\{i \cdot t \cdot [(E_n - E_m)/\hbar + \Omega]\}, \; V^{(k)}{}_{nm} = \int \varphi_n^* \cdot \hat{U}_k \cdot \varphi_m dq, \; \Omega = \{\omega_0, \omega_1\}. \qquad (14)$$

Рассмотрим динамику трехуровневой системы ($|0\rangle, |1\rangle, |2\rangle$). Будем считать, что частота внешнего возмущения и собственные значения энергий этих уровней удовлетворяют соотношениям:

$$m=1, n=0, \quad \hbar\omega_0 = E_1 - E_0; \quad m=2, n=0 \quad \hbar(\omega_0 + \omega_1) = E_2 - E_0, \quad \hbar\omega_1 = E_2 - E_1. \qquad (15)$$

Соотношения (15) указывают на тот факт, что частота $\omega_0$ внешнего возмущения является резонансной для переходов между нулевым и первым уровнями, а частота $\omega_1$ является резонансной для переходов между первым и вторым уровнями. Используя эти соотношения в системе (13), можно ограничиться тремя уравнениями:

$$i\dot{A}_0 = A_1, \quad i\dot{A}_1 = A_0 + \mu A_2, \quad i\dot{A}_2 = \mu A_1. \qquad (16)$$

Систему уравнений (16) представим в несколько другом виде:

$$\ddot{A}_1 + \Omega^2 A_1 = 0, \quad i\dot{A}_0 = A_1, \quad i\dot{A}_2 = \mu A_1, \qquad (16а)$$

где $\Omega^2 = (1 + \mu^2)$

Схема энергетических уровней при этом представлена на рисунке 1.

В (16) для простоты и удобства положили $V_{12} = V_{21}$; $V_{10} = V_{01}$; $\dot{A}_i = dA_i/d\tau$, $\tau = V_{10} \cdot t/\hbar$. Кроме того, введен параметр $\mu \equiv V_{12}/V_{10}$.

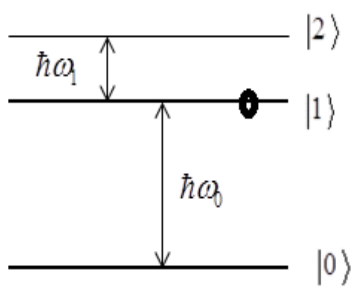 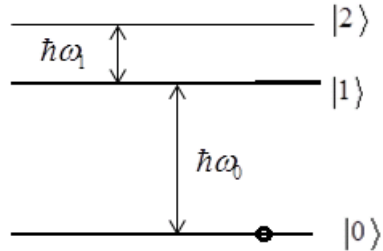 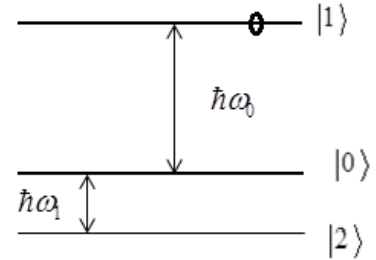

*Рис.1а. Схема энергетических уровней. В начальный момент времени система находилась на возбужденном уровне $|1\rangle$*

*Рис.1b. Схема энергетических уровней. При $\tau = 0$ система находилась на стационарном, невозбужденном уровне $|0\rangle$*

*Рис.1с. Схема энергетических уровней. При $\tau = 0$ система находилась на возбужденном уровне $|1\rangle$*

Пусть, в начальный момент времени ($t = 0$) рассматриваемая квантовая система находится на первом, возбужденном уровне. Тогда, как легко видеть, решениями системы (16) будут функции:

$$A_0 = \frac{1}{i \cdot \Omega} \sin(\Omega \cdot t), \ A_1 = \cos(\Omega \cdot t), \ A_2 = -i \sin(\Omega \cdot t). \quad (17)$$

Из решения (17) следует, что чем больше будет параметр $\mu$, тем меньше будет вероятность, что система из возбужденного состояния перейдет в невозбужденное, стационарное состояние. Следует несколько слов сказать о параметре $\mu$. Физически этот параметр определяет отношение числа квантов низкочастотного возмущения, которое ответственно за переходы между первым и вторым уровнями к числу квантов высокочастотного возмущения, которое определяет переходы между первым и нулевым уровнями. Чем больше будет это отношение, тем меньше будет вероятность того, что возбужденная система перейдет в невозбужденное состояние.

Обратим внимание, что система уравнений (16) третьего порядка, а система (16а) – четвертого. В результате, если мы будем решать задачу с начальными условиями $|A_0(0)|^2 = 1, |A_1(0)| = |A_2(0)| = 0$, т.е. в начальный момент времени система находится в основном невозбужденном уровне, а быструю динамику совершают те уровни, на которые система из исходного уровня должна перейти под влиянием внешнего ВЧ-возмущения. В этом случае запишем решение системы (16) в виде:

$$A_1 = a \cdot \exp(i \cdot \Omega \cdot t) + b \cdot \exp(-i \cdot \Omega \cdot t) .$$

Учтем начальные условия для $A_1$ ($A_1(0) = 0$). Тогда решениями системы уравнений (16) будут функции:

$$A_1 = a \cdot \left[\exp(i \cdot \Omega \cdot t) - \exp(-i \cdot \Omega \cdot t)\right], \ A_0 = -\frac{a}{\Omega} \cdot \left[\exp(i \cdot \Omega \cdot t) + \exp(-i \cdot \Omega \cdot t)\right] + C_0,$$

$$A_2 = -\frac{a \cdot \mu}{\Omega} \cdot \left[\exp(i \cdot \Omega \cdot t) + \exp(-i \cdot \Omega \cdot t)\right] + C_2.$$

У нас оказалось три константы $a, C_0, C_2$ и только два не использованных начальных условия. Такое положение возникло из-за того, что второе уравнение первого порядка в системе (16) заменили первым уравнением системы (16а), которое является уравнением второго порядка. Поэтому кроме начальных условий необходимо, чтобы полученные решения удовлетворяли еще и второму уравнению системы (16). Из начальных условий получим:

$$C_0 = 1 + a/\Omega \qquad C_2 = (a \cdot \mu)/\Omega.$$

Потребуем, чтобы полученные решения удовлетворяли уравнению $i\dot{A}_1 = A_0 + \mu A_2$.

Отсюда находим следующую связь между постоянными $C_0$ и $C_2$: $C_0 + \mu \cdot C_2 = 0$ или $1 + a/\Omega + (a \cdot \mu)/\Omega$. В результате, находим значение постоянной $a$: $a = -1/\Omega$.

Окончательно выражения для амплитуд волновых функций приобретают вид:

$$A_1 = -(2i/\Omega)\sin(\Omega t), \ A_0 = 1 - \frac{1}{\Omega^2}[1 - \cos(\Omega t)], \ A_2 = \frac{2\mu}{\Omega^2}[1 - \cos(\Omega t)] . \quad (17а)$$

Из вида решений (17а) следует важный и несколько неожиданный результат. Он заключается в том, что если параметр $\mu$ будет большим, то, несмотря на тот факт, что внешнее стабилизирующее воздействие не оказывает влияния на основное состояние системы, однако это состояние оказывается устойчивым. Таким образом, оказывается возможным стабилизировать неустойчивые состояния квантовых систем не только, действуя на те состояния, в которых находится квантовая система, а, действуя только на те состояния (делая их динамическими), в которые система должна перейти.

Приведем еще один важный случай, который подтверждает такую возможность стабилизации. Это случай, когда квантовая система находится на возбужденном энергетическом уровне $|1\rangle$ и под действием возмущения $\hbar\omega_0 = E_1 - E_0$ должна перейти на стационарный уровень $|0\rangle$ (см. рисунок 1с). Кроме возмущения на частоте имеется стабилизирующее возмущение на частоте $\hbar\omega_1 = E_0 - E_2$. Система уравнений, которая описывает такую квантовую систему, имеет вид:

$$i\dot{A}_1 = A_0, \quad i\dot{A}_0 = A_1 + \mu A_2, \quad i\dot{A}_2 = \mu A_0.$$

$$\text{or} \quad \ddot{A}_0 + \Omega^2 A_0 = 0, \quad i\dot{A}_1 = A_0, \quad i\dot{A}_2 = \mu A_0 \qquad (6b)$$

Решением системы (6b) будут функции:

$$A_0 = \frac{1}{\Omega}\sin(\Omega \cdot t); \quad A_1 = 1 - \frac{1}{i\cdot\Omega^2}\cos(\Omega \cdot t); \quad A_2 = \frac{\mu}{i\cdot\Omega^2}\left[1 - \cos(\Omega \cdot t)\right] \qquad (17b)$$

Из этого решения следует, что как только параметр $\mu \equiv V_{02}/V_{10} \gg 1$ становится достаточно большим, то вероятность квантовой системы перейти из возбужденного состояния $|1\rangle$ в основное стационарное состояние $|0\rangle$ стремится к нулю.

## СРАВНЕНИЕ МЕХАНИЗМА СТАБИЛИЗАЦИИ НЕУСТОЙЧИВЫХ СИСТЕМ В БЫСТРООСЦИЛЛИРУЮЩЕМ ПОЛЕ С ПРИНЦИПОМ ЮЛЫ
### Общие соображения

Выскажем вначале кратко некоторые общие соображения, которые позволяют понять механизм стабилизации, который мы называем механизмом стабилизации юлы. В подавляющем большинстве случаев неустойчивые состояния динамических систем локально характеризуются особыми точками типа "седло". Неустойчивые узлы и фокусы встречаются значительно реже. Поясним на примере неустойчивой точки типа "седло", как такая особая точка может быть трансформирована в эллиптическую точку (в точку типа "центр"). Фазовые портреты окрестности седловой точки представлены на рисунках 2-4. Уравнения на фазовой плоскости, которые описывают динамику фазовых траекторий в окрестности седловой точки, имеют вид:

$$\dot{x}_0 = \gamma \cdot x_1 \qquad \dot{x}_1 = \gamma x_0.$$

(18)

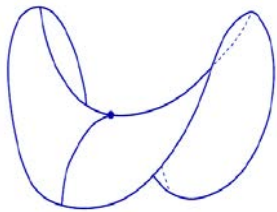 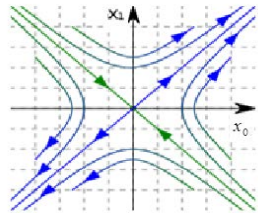 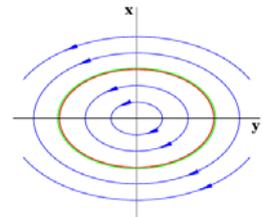

*Рис.2. Фазовый портрет в окрестности особой точки типа "седло"*

*Рис.3. Фазовый портрет в окрестности особой точки типа "седло"*

*Рис.4. Фазовый портрет в окрестности особой точки типа "центр"*

В большинстве реальных случаев каждая из зависимых переменных уравнений (18) представляет собой какую-то характеристику собственной степени свободы изучаемой динамической системы. Например, это могут быть комплексные амплитуды нелинейно-взаимодействующих волн. Поэтому в этой работе будем считать, что каждое такое уравнение первого порядка описывает одну из степеней свободы изучаемых систем. Пусть, теперь перед нами стоит задача, чтобы окрестность седловой особой точки преобразовать в окрестность, которая соответствует устойчивой особой точке, например, эллиптической точке (рис. 4). Для этого можно поступить таким образом. Введем в нашу изучаемую систему дополнительную степень свободы, которая связана с одной из степеней свободы изучаемой неустойчивой системы. Простейшая модель, которая будет описывать динамику системы в окрестности седловой точки при такой модификации системы, будет отличаться от уравнения (18) только добавлением еще одного уравнения:

$$\dot{x}_0 = \gamma \cdot x_1 + \delta \cdot x_2; \qquad \dot{x}_1 = \gamma x_0; \qquad \dot{x}_2 = -\delta \cdot x_0. \qquad (19)$$

Здесь в отличие от (18) мы ввели дополнительную степень свободы. Причем, эта новая степень свободы связана с одной из степеней свободы исходной системы коэффициентом $\delta$.

Система уравнения (19) эквивалентна уравнению линейного маятника:

$$\ddot{x}_0 + \left(\delta^2 - \gamma^2\right) x_0 = 0 \qquad (20)$$

Из уравнения (20) сразу видно, что как только коэффициент, описывающий связь между степенями свободы, будет большим, чем инкремент неустойчивости ($\delta > \gamma$), неустойчивая седловая точка превратится в эллиптическую точку. Фазовое пространство, представленное на рис. 3, станет фазовым пространством, представленном на рис. 4.

Этот простой алгоритм перевода неустойчивой седловой точки в эллиптическую точку хорошо характеризует принцип юлы. Действительно, если мы не вводили дополнительную степень свободы, то наша система была неустойчивой (юла падает). Причем, время развития неустойчивости ($T \sim 1/\gamma$) можно отождествить со временем падения юлы. Включение дополнительной степени свободы, которая стабилизирует нашу систему, аналогично наличию вращения юлы. Более того, в образе юлы имеются не только качественные аналогии, но и количественные. Действительно, чтобы вертикальное положение юлы было устойчивым, необходимо, чтобы период вращения был меньшим времени падения. В наших моделях (см. уравнение (20)), чтобы система стала устойчивой нам также необходимо, чтобы коэффициент связи был большим инкремента неустойчивости $\delta > \gamma$. Более того, если инкремент неустойчивости равен нулю, то система уравнений (19) или (20) описывает просто колебания с частотой $\delta = 2\pi / T_{rot}$. Таким образом, имеется качественная и количественная аналогия рассматриваемого механизма стабилизации с механизмом стабилизации вертикального положения юлы.

Сделаем следующее замечание. Мы привыкли к тому, что увеличение числа степеней свободы изучаемой динамической системы приводит к более жестким условиям для реализации ее устойчивого состояния. Действительно, пусть наша физическая система описывается следующей системой уравнений:

$$\dot{Z}_n = F_n(\vec{Z}, t) . \qquad (21)$$

Характер устойчивости этой системы в выбранной точке фазового пространства $\vec{Z}_0$ описывается линейной системой уравнений, которая описывает динамику малых отклонений $\vec{x} = \vec{Z} - \vec{Z}_0$:

$$\dot{\vec{x}} = \hat{A}\vec{x} . \qquad (22)$$

Во многих случаях коэффициенты матрицы $\hat{A}$ можно считать постоянными величинами. Тогда для определения характера особой точки, в которой написана система (22), мы должны найти корни характеристического уравнения:

$$\det(\hat{A} - \lambda \cdot \hat{I}) = 0; \qquad \alpha_0 \lambda^n + \alpha_1 \lambda^{n-1} + \alpha_2 \lambda^{n-2} + ... + \alpha_{n-1} \lambda + \alpha_n = 0 . \qquad (23)$$

Критерий Рауса-Гурвица утверждает, что чем выше порядок рассматриваемых уравнений (22) и (23), тем труднее удовлетворить условиям реализации устойчивой динамики этой системы. В приведенном нами выше примере мы увеличили число степеней свободы. Но добились прямо противоположного результата. Может показаться, что это исключительный случай. Однако это не так. Ниже и более подробно в работах [3-8] было показано, что введение дополнительной степени свободы в значительно более сложных системах таких, например, которые описывают стабилизацию потоков излучения в плазме, введение такой дополнительной степени свободы также могло приводить к стабилизации неустойчивых состояний. Этот результат можно понять если принять во внимание, что добавочная степень свободы (стабилизирующая) порождает новую быструю динамику. Наличие этой быстрой динамики порождает иерархию процессов и конкуренцию [8,9]. При этом одна из степеней свободы исходной (невозмущенной) системы "выпадает" из медленной динамики.

**Сравнение принципа юлы с маятником Капицы**

Рассмотренные выше квантовый эффект Зенона и принцип юлы, содержат в себе основной элемент, который заключается в том, что происходят быстрые изменения каких-то характеристик стабилизируемых систем. Эта особенность напоминает особенности, которые характерны для динамики частиц в быстроосциллирующем поле. Наиболее простой алгоритм описания движения систем в таких полях был предложен Капицей при анализе динамики математического маятника, точка подвеса которого быстро осциллирует [10,11]. Этот пример содержит наиболее важные особенности динамики систем в быстроосциллирующем поле.

Воспользуемся этой простой моделью для сравнения механизма стабилизации динамики систем в быстроосциллирующем поле и принципом юлы. Найдем характеристики, которые являются похожими, а также те характеристики, которые отличают эти два механизма стабилизации.

Прежде всего, опишем кратко динамику математического маятника, параметры которого претерпевают быстрые осцилляции. При описании такой динамики будем пользоваться алгоритмом, который описан в книге Ландау [12]. Уравнение, которое описывает такой математический маятник, имеет вид:

$$\ddot{x} + \left(\Omega^2 + \varepsilon \cos(\omega \cdot t)\right) \sin x = 0 . \tag{24}$$

В уравнении (24) $\Omega$ - собственная частота малых колебаний маятника; $\omega$ - частота быстрых осцилляций параметров маятника. В частности, эта частота может быть частотой изменения положения точки подвеса математического маятника. Предполагается, что эта частота значительно больше собственной частоты маятника ($\omega \gg \Omega = 2\pi/T$). Для дальнейшего, удобно перейти к новой независимой переменной $\tau = \Omega t$. Тогда уравнение (24) можно переписать:

$$\ddot{x} + \left(1 + q \cos(\omega_N \cdot \tau)\right) \sin x = 0, \tag{25}$$

где - $q = \varepsilon / \Omega^2$, $\omega_N = \omega / \Omega \gg 1$.

Следуя [12], уравнение (25) запишем в виде:

$$\ddot{x} = -\frac{dU}{dx} + f(x,t), \tag{26}$$

где $\dfrac{dU}{dx} = \sin x \quad f(x,t) = -q \cdot \cos \omega_N \tau \cdot \sin x$.

Далее представим зависимую переменную в виде суммы медленноменяющейся ($X(t)$) и быстроменяющейся ($\xi(t)$) величины: $x(t) = X(t) + \xi(t)$. Подставим это выражение в уравнение (26). Будем считать, что быстроменяющаяся величина мала по сравнению с медленноменяющейся величиной. Разложим функции, входящие в правую часть уравнения (26) в ряд Тейлора в окрестности функции $X(t)$. Ограничиваясь первыми неисчезающими членами этого разложения, уравнение (26) можно переписать в виде:

$$\ddot{X}(t) + \ddot{\xi}(t) = -\left[\Omega^2 \sin X + \varepsilon \xi \cos X \cdot \cos \omega t\right] - \varepsilon \sin X \cdot \cos \omega t - \Omega^2 \cos X \cdot \xi. \tag{27}$$

Применим к левой и правой частям уравнения (27) процедуру усреднения по быстроменяющейся величине, т.е. проинтегрируем эти части по периоду $\tau = 2\pi/\omega$:

$$\langle Z \rangle = \frac{1}{\tau}\int_0^\tau Z \cdot dt.$$

Учитывая, что медленноменяющиеся величины "не замечают" такого усреднения, найдем следующее выражение для быстроменяющейся величины:

$$\xi = -\left(q/\omega_N^2\right)\sin X \cdot \cos \omega_N \tau \tag{28}$$

и уравнение, которое описывает медленную динамику маятника.

$$\ddot{X} = -\frac{dU_{eff}}{dX}, \tag{29}$$

где - $U_{eff} = -\cos X + \alpha \sin^2 X$, $\alpha = q^2/4\omega_N^2$.

Устойчивое положение математического маятника будет соответствовать минимуму эффективного потенциала, т.е. определяется равенством нулю производной от потенциала: $\partial U/\partial x = \sin x \cdot [1 + 2\alpha \cdot \cos x] = 0$. Видно, что нижнее положение маятника ($x=0$) всегда устойчиво. Устойчивыми будут также и все те положения маятника, для которых $[1 + 2\alpha \cdot \cos x] = 0$. В частности, верхнее (вертикальное) положение маятника ($x=\pi$) будет устойчивым при выполнении условия $q^2 > 2\omega_N^2 \gg 1$. На рис. 6-7 представлен вид потенциала $U(x) = -\cos(x) + \alpha \cdot \sin^2(x)$ при значениях $\alpha = 0,5$ (рис.6) и при $\alpha = 1,2$ (рис. 7).

Из этих рисунков видно, что при малых значениях параметра $\alpha$ эффективный потенциал содержит только одну точку минимума ($x=0$), которая соответствует нижнему положению маятника. Все остальные положения маятника неустойчивы. Однако, начиная с величины $\alpha > 1$, в точке $x = \pi$ появляется локальный минимум. Глубина этого локального минимума растет с увеличением параметра $\alpha$. Растут при этом и степень устойчивости и область устойчивых значений угловой переменной $x$.

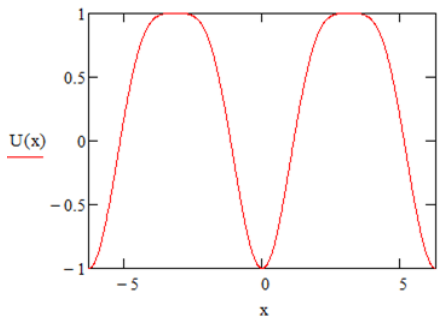
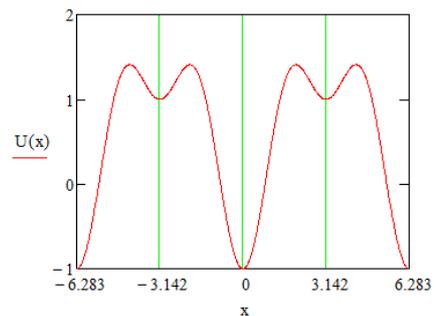

*Рис. 6. Эффективный потенциал при $\alpha = 0,5$*  *Рис. 7. Вид эффективного потенциала при $\alpha = 1,2$*

Для сравнения динамики маятника Капицы с динамикой системы, которая стабилизируется с использованием принципа юлы удобно непосредственно рассмотреть динамику математического маятника в окрестности неустойчивой точки $x = \pi$. Динамику маятника в окрестности этой точки можно описать системой уравнений:

$$\ddot{x}_0 - \Omega^2 x_0 = \varepsilon \cos(\omega \cdot t) x_0 = x_1 \cdot x_0 \tag{30}$$
$$\ddot{x}_1 + \omega^2 x_1 = 0, \quad x_1(0) = \varepsilon;\; \dot{x}_1(0) = 0.$$

Здесь $x_0 = x - \pi$.

Второе уравнение – уравнение для $x_1$ - описывает динамику внешней стабилизирующей силы. Обратим внимание на то, что динамика этой силы не зависит от динамики стабилизируемой системы (от $x_0$). Таким образом, в данном случае внешнее воздействие является не самосогласованным с динамикой маятника. Эта особенность характерна для стабилизации и других систем при использовании для стабилизации внешних быстроосциллирующих сил. Во многих случаях эта особенность является полезной.

Перепишем уравнение (30) в тех же переменных, что и уравнение (25):
$$\ddot{x}_0 - x_0 = q \cdot \cos(\omega_N \cdot \tau) x_0 = x_0 x_1 \equiv f(x_0, \tau) \tag{31}$$
$$\ddot{x}_1 + \omega^2 x_1 = 0, \quad x_1(0) = q; \; \dot{x}_1(0) = 0.$$

Систему уравнений (31) будем рассматривать как систему, которая описывает движение в постоянном (во времени) потенциале ($U(x) = -x^2/2$) и под влиянием внешней быстроосциллирующей силы, которая меняет параметры маятника ( $q\cos(\omega \cdot t) x_0 \equiv f(x_0, t)$). Тогда для анализа такого движения можно использовать описанный выше алгоритм исследования. В результате для описания медленной динамики маятника получим уравнение, которое по форме совпадает с уравнением (29):
$$\ddot{X} = -\frac{dU_{eff}}{dX} \quad . \tag{32}$$

Здесь $U_{eff}(x) = -\frac{x^2}{2} + \frac{1}{2\omega_N^2}\langle f^2 \rangle = -\frac{x^2}{2}\left[1 - \frac{q^2}{2\omega_N^2}\right]$.

Устойчивое состояние соответствует минимуму этого потенциала:
$$q^2 > 2\omega_N^2 \gg 1 \quad . \tag{33}$$

Естественно, что условие (33) совпадает с полученным выше условием устойчивости вертикального положения маятника. В качестве конкретного примера рассмотрим маятник, длина которого равна $l$, и который колеблется в поле тяжести $g$, а точка подвеса которого меняется с частотой $\omega \gg \Omega = \sqrt{l/g}$. Если при этом максимальное отклонение точки подвеса равно $a$, то условием устойчивости вертикального положения маятника будет:
$$\omega > \sqrt{2} \cdot \Omega \cdot \frac{l}{a}; \quad \Omega = \sqrt{l/g} \quad . \tag{34}$$

По определению (по постановке задачи) это большая величина. Если смотреть только на левую часть уравнения (31), то это уравнение описывает неустойчивую стационарную точку типа "седло". Наличие быстроосциллирующей параметрической силы (правая часть уравнения (31)), как мы видели, при условии (33) приводит к преобразованию седловой точки к устойчивой точке типа "центр". Такой сценарий стабилизации вертикального положения математического маятника удобен для сравнения его со сценарием стабилизации при использовании принципа юлы. Действительно, рассмотрим неустойчивую седловую точку, которая соответствует уравнению (31) в отсутствии правой стабилизирующей силы:
$$\dot{x}_0 = \Omega \cdot x_1 \quad \dot{x}_1 = \Omega \cdot x_0 \quad . \tag{35}$$

Предположим теперь, что переменная $x_0$ связана линейной связью с некоторой другой переменной $x_2$. Система уравнений, которая описывает такую модифицированную систему, может иметь вид:
$$\dot{x}_0 = \Omega \cdot x_1 + \delta \cdot x_2$$
$$\dot{x}_1 = \Omega \cdot x_0, \quad \dot{x}_2 = -\delta \cdot x_0 \quad . \tag{36}$$

Система уравнений (36) эквивалентна уравнению линейного маятника:
$$\ddot{x}_0 + \left(\delta^2 - \Omega^2\right)x_0 = 0 \quad . \tag{37}$$

Из этого уравнения следует, что как только выполняется условие $\delta > \Omega$, то неустойчивая седловая точка становится устойчивой точкой типа "центр". Сравнивая это условие стабилизации с условием стабилизации вертикального положения маятника (33), видим, что использование принципа юлы для преобразования неустойчивой седловой точки в устойчивую точку значительно более эффективно. Следует, конечно, иметь в виду, что если речь идет об обычном математическом маятнике, который колеблется в гравитационном поле, то использование принципа юлы для стабилизации вертикального положения такого маятника физически выглядит не слишком удобным. Однако модель математического маятника является одной из наиболее распространенной, наиболее удобной моделью, к которой сводится описание динамики большого числа разнообразных физических систем. Поэтому во всех тех случаях, когда принцип юлы может быть использован - он оказывается эффективнее простого параметрического воздействия на параметры неустойчивой системы. Отметим еще одно различие в механизме стабилизации маятника Капицы и при использовании принципа юлы. Это различие легко увидеть, если сравнивать систему уравнений (31) и (36). В первом случае мы видим, что внешнее стабилизирующее возмущение является независимым параметром. Колебания самого стабилизируемого маятника не влияют на динамику этой внешней силы. В отличие от этого стабилизирующее воздействие в системе (36) является самосогласованным. Колебания стабилизируемой системы существенно влияют на динамику самой стабилизируемой силы.

## ПРИМЕРЫ ИСПОЛЬЗОВАНИЯ ПРИНЦИПА ЮЛЫ ДЛЯ СТАБИЛИЗАЦИИ НЕУСТОЙЧИВЫХ СИСТЕМ

В этом разделе кратко опишем некоторые примеры использования принципа юлы. Более полное изложение описанных примеров можно найти в работах [3-8].

### Подавление СИ

В работах [3,7] (смотри также [5,8]) было показано, что принцип юлы может быть использован для подавления синхротронного излучения (СИ), т.е. для стабилизации высоких уровней Ландау. Ниже мы опишем условия необходимые для подавления СИ. Особый интерес при этом представляет сравнение классического эффекта подавления с квантовым эффектом подавления.

**Классическая оценка условий подавления СИ**. Прежде всего, рассмотрим, как может быть подавлено СИ в рамках классической электродинамики. При этом будем считать, что если под действием внешней электромагнитной волны электрон будет выходить за пределы угла излучения СИ, то его излучение будет подавлено. Для оценки необходимой напряженности поля воспользуемся тем фактом, что длину формирования СИ можно оценить величиной $l \approx \lambda \gamma^2$ [13]. Здесь $\lambda$ - длина излученной волны, $\gamma$ - энергия частицы. Соответствующий угол излучения будет равен $\theta \approx 1/\gamma$. Время, за которое частица пройдет путь равный длине формирования, будет равно $\tau \approx l/c = \lambda \gamma^2 / c$. Для того, чтобы поле внешней стабилизирующей волны "выбивало" частицу из угла излучения, необходимо, чтобы частота этой волны была больше величины $\Omega > 2\pi/\tau = 2\pi c / \lambda \gamma^2$. С другой стороны угол, под которым движется частица, можно оценить величиной $\theta \approx r_\perp / r_\| = r_\perp / l \ll 1$. Поперечное отклонение частицы в поле внешней электромагнитной волны оценим величиной $r_\perp \approx \dfrac{eE}{m_\perp \Omega^2}$. Чтобы было подавление процесса излучения необходимо, чтобы частица вышла (под действием возмущения) за пределы конуса излучения: $r_\perp / l > 1/\gamma$. Из этого условия можно определить необходимую напряженность поля: $(eE / m_0 c\Omega) > 2\pi$.

**Квантовая оценка условий подавления СИ**. Оценим теперь, необходимую для стабилизации, напряженность электрического поля внешней электромагнитной волны, если будем учитывать квантовый эффект Зенона. В соответствии с принципом юлы первым шагом для определения условий подавления является определение времени жизни возбужденного состояния. Это время для СИ в синхротроне в отсутствии возмущения можно оценить формулой $T_L = \dfrac{\hbar \cdot R}{r_0 \cdot mc^2 \cdot \gamma}$ [14]. Здесь $r_0$ – классический радиус электрона; $R$ – радиус орбиты электрона в синхротроне. Если в качестве примера обычные параметры синхротрона: $R = 100 cm$, $E = mc^2 \cdot \gamma = 500 MeV$, то время жизни окажется порядка $10^{-9}$ sec. Вторым шагом является нахождение частоты Раби. В уравнение Шредингера (Дирака) входит величина потенциала внешней волны. Этот потенциал оценим величиной $V \approx eE\lambda$. Соответственно, частота Раби будет равной $\Omega_R = V / \hbar = eE\lambda / \hbar$.

Мы видели, что для проявления эффекта подавления необходимо, чтобы эта частота была большей чем $2\pi / T_{LF}$ ($\Omega_R >> 2\pi / T_{LF}$). Отсюда можно получить следующую оценку на величину напряженности поля, необходимой для подавления СИ: $E > \dfrac{10^{10} \hbar}{e\lambda} 300 \sim \dfrac{1}{\lambda} 10^{-5} (V/cm)$. Из этой оценки видно, что в квантовом случае величина напряженности поля, необходимая для стабилизации, на много порядков меньше, чем напряженность поля, полученная в рамках классической электродинамики. Этот результат легко объясняется тем фактом, что в квантово-механическом рассмотрении для подавления необходимо только, чтобы частота Раби была выше, чем обратное время жизни электронов в возбужденном состоянии. В рамках классической электродинамики такие процессы просто отсутствуют.

Описанный механизм (принцип юлы) может с успехом быть использован для стабилизации классических систем. Ниже это будет показано на примере подавления плазменно-пучковой неустойчивости и на примере подавления распадной неустойчивости при распространении потоков излучения в нелинейных средах, в частности, в плазме.

### Подавление плазменно-пучковой неустойчивости

Пусть, у нас имеется плазменный цилиндр ($0 < r < R_p$). Он помещен в металлический кожух того же радиуса. Плазма помещена в сильное внешнее магнитное поле. Пучок проходит вдоль оси плазменного цилиндра. Радиус пучка совпадает с радиусом плазмы. В металлическом кожухе имеются элементы связи с внешней электродинамической структурой (например, отверстия (щели)). В качестве внешней электродинамической структуры может быть выбрана спираль радиуса $R_H$. Таким образом, у нас имеется три основных колебательных системы: это плазма ($n_p$), пучок ($n$) и внешняя колебательная структура ($E_2$). Для эффективного взаимодействия колебаний в плазме и колебаний во внешней электродинамической системе их частоты должны совпадать ($\omega_p = |k_\perp| c$). Должны совпадать и продольные волновые числа. Система уравнений, которая описывает динамику такой колебательной системы, может быть представлена в виде системы трех связанных осцилляторов:

$$\ddot{n}_p + \omega_p^2 n_p = -\omega_p^2 n + i\mu \frac{k_z n_{0p} e}{m} E_2 \qquad (38)$$

$$\ddot{n} + \omega_b^2 n - 2ik_z \dot{n} - k_z^2 V^2 n = -\omega_b^2 n_p + i\mu \frac{k_z n_0 \cdot e}{m} E_2$$

$$\ddot{E}_2 - k_\perp^2 c^2 E_2 = \mu_1 \frac{4\pi e}{ik_z}(n_p + n).$$

Здесь $\mu_1 = \mu / G$, $G$ - норма поля волны во внешней структуре, $k_\perp^2 = \left(\lambda_n^2 / R_H - k_z^2\right)$, $E_2$ - продольная компонента электрического поля волны во внешней электродинамической структуре $\lambda_n$ - корни функций Бесселя ($J_0(\lambda_n) = 0$), $\mu$ - коэффициент связи плазменной волны с собственной волной внешней электродинамической структуры.

Из уравнений (38) можно получить следующее дисперсионное уравнение:

$$\left[1 - \frac{\omega_p^2}{\omega^2} - \frac{\omega_b^2}{(\omega^*)^2}\right] - \frac{\mu\mu_1}{(\omega^2 + k_\perp^2 c^2)}\left(\frac{\omega_p^2}{\omega^2} + \frac{\omega_b^2}{(\omega^*)^2}\right) = 0. \tag{39}$$

Из этого дисперсионного уравнения видно, что в отсутствии связи между полями в плазме и во внешней структуре ($\mu_k = 0$) получается обычное дисперсионное уравнение системы плазма-пучок. Наоборот, если пучок отсутствует ($\omega_b^2 = 0$), то оно переходит в дисперсионное уравнение, которое описывает перекачку энергии между плазменными волнами и волнами внешней структуры. Частота такой перекачки равна $\Omega = \sqrt{\mu\mu_1}/2$. В соответствии с принципом юлы можно рассчитывать, что когда эта частота окажется больше инкремента пучковой неустойчивости, то такая неустойчивость будет подавлена. Аналитические и численные исследования показали, что как только выполняется неравенство $\sqrt{\mu\mu_1}/2 > \left(\omega_b^2 \omega_p / 2\right)^{1/3}$ плазменно-пучковая неустойчивость не развивается.

### Стабилизация потоков излучения в плазме

При распространении волн в плазме развивается неустойчивость (распадная неустойчивость). Она может быть полезной. Однако она может быть и нежелательной, вредной. Особенно в том случае, когда эта неустойчивость переходит в стохастический режим [15-16]. В этом случае с нею следует бороться. Ниже покажем, что такие неустойчивости могут быть подавлены. Для этого достаточно, чтобы одна из участвующих в трехволновом взаимодействии волн участвовала в некотором дополнительном периодическом процессе (стабилизирующем процессе). Простейшая система уравнений, которая описывает такие процессы, может быть представлена в следующем виде:

$$\frac{dA_0}{dt} = -VA_1 A_2 + \frac{\mu}{2i} A_3 \; ; \qquad \frac{dA_3}{dt} = \frac{\mu}{2i} A_0 \; ; \tag{40}$$

$$\frac{dA_1}{dt} = VA_0 A_2^* \; ; \qquad \frac{dA_2}{dt} = VA_1^* A_0.$$

Эта система уравнений описывает взаимодействие четырех волн. Причем, две из них нулевая и третья в наших обозначениях связаны друг с другом линейной связью. Связь характеризуется коэффициентом связи $\mu$. Если другие волны отсутствуют, то происходит периодическая перекачка энергии из основной волны в стабилизирующую (третью) волну и обратно. Частота такой перекачки равна $\Omega = \mu/2$. Три волны (нулевая, первая и вторая) взаимодействуют через нелинейность. Если коэффициент связи равен нулю ($\mu = 0$), то система (40) описывает обыкновенное трехволновое взаимодействие волн, динамика которого хорошо изучена (см., например, [17,18]). Инкремент распадной неустойчивости равен $\delta = V|A_0(0)|$. Отметим, что если знак первого члена правой части первого уравнения с минуса поменять на плюс, то такая система будет описывать взрывную неустойчивость, которая также детально изучена.

При включении стабилизирующей волны при выполнении условия $\mu/2V > |A_0(0)|$ во всех случаях наблюдался процесс стабилизации распадной неустойчивости.

### Подавление локальной неустойчивости

Ниже будет видно, что принцип юлы может быть полезным для решения проблемы подавления режимов с динамическим хаосом. Причиной возникновения режимов с динамическим хаосом является локальная неустойчивость. При этом в фазовом пространстве имеется большое число седловых точек. В частности, в гомоклинических структурах таких точек бесконечно много. В окрестности каждой такой точки близко расположенные траектории экспоненциально разбегаются друг от друга. Выше мы видели, что если одна из степеней свободы, которая участвует в нелинейном взаимодействии, окажется задействованной в некотором добавочном быстром процессе, то неустойчивая седловая точка может быть преобразована в устойчивую точку.

Можно ожидать, что аналогичное преобразование седловых точек можно реализовать в системах с динамическим хаосом. Для этого нужно организовать связь, по крайней мере, одной из зависимых переменных с некоторой дополнительной зависимой переменной. Ниже мы увидим, что действительно такой алгоритм может быть реализован.

В качестве примера рассмотрим несколько измененную модель Лоренца:

$$\dot{x} = \sigma(y-z) - \mu w, \quad \dot{y} = r \cdot x - y - x \cdot z, \quad \dot{z} = x \cdot y - b \cdot z, \quad \dot{w} = \mu x. \quad (41)$$

Если коэффициент связи равен нулю ($\mu = 0$), то система уравнений (41) описывает известную модель Лоренца. При значениях параметров $\sigma = 10$, $b = 8/3$, $r = 28$ эта система находится в режиме с динамическим хаосом. Динамика такой системы изучена очень подробно. Можно ожидать, что если любая из компонент системы Лоренца ($x; y; z$) окажется связанной с какой-то четвертой компонентой ($w$), и связь между этими компонентами будет такой, что период обмена энергиями между этими компонентами будет меньшим, чем обратный инкремент локальной неустойчивости, то динамика системы Лоренца будет сложной, но регулярной. Система (41) исследовалась численными методами. Ниже приведены некоторые результаты этого исследования. На рис. 7,8 представлена динамика классической системы Лоренца (без связи с внешним дополнительным компонентом).

Из рисунков видно, что наблюдается привычная динамика системы Лоренца: спектр этой динамики широкий и корреляционная функция быстро спадает. Рассмотрим теперь ситуацию, когда внешняя динамическая переменная связана с первым компонентом системы Лоренца (41). Численные расчеты показывают, что увеличение коэффициента связи до величины порядка 5 мало сказывается на статистических характеристиках системы Лоренца. Однако, начиная где-то с пяти, шести, эта динамика становится регулярной. Эти утверждения иллюстрируются ниже приведенными рисунками.

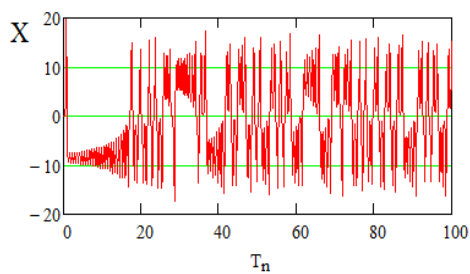 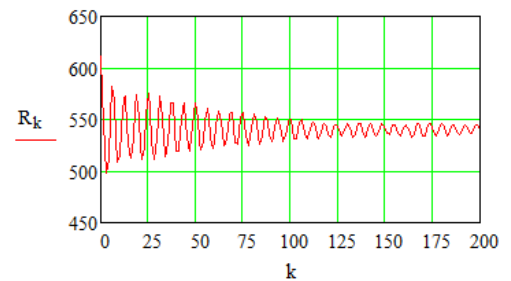

*Рис. 7. Обычная динамика переменной $x$ системы Лоренца без влияния стабилизирующей переменной $\mu = 0$*

*Рис. 8. Автокорреляционная функция переменной $x$ при $\mu = 0$*

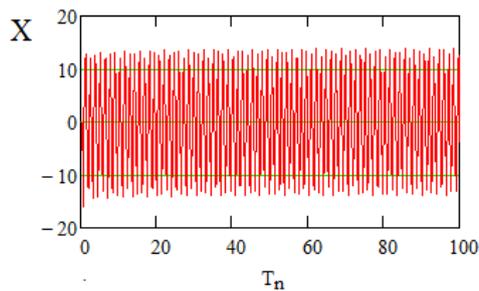 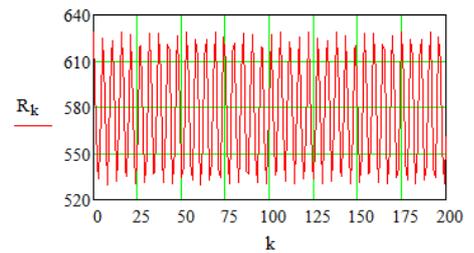

*Рис. 9. Динамика переменной $x$ системы Лоренца при наличии стабилизирующей переменной ($\mu = 6$)*  *Рис. 10. Автокорреляционная функция переменной $x$ при $\mu = 6$*

Из рисунков видно, что динамика стала значительно более регулярной. Амплитуда корреляционной функции практически не меняется. Аналогичные результаты получаются и при связи внешнего стабилизирующего компонента с двумя другими компонентами ($y, z$) системы Лоренца. Во всех случаях при достаточно большой величине связи динамика системы Лоренца становится регулярной.

Следует отметить, что к системе (41) при $\mu = 0$ приводится большое количество разнообразных физических систем. Поэтому возникает вопрос о физической природе внешнего компонента, который может стабилизировать систему Лоренца. Ясно, что во всех конкретных случаях это будут разные физические переменные. В частности, укажем, что если система Лоренца описывает динамику одномодового лазера, то зависимая переменная $x$ определяет амплитуду поля лазера. В этом случае легко себе представить практическую реализацию механизмов подавления. Достаточно связать поле одномодового лазера с некоторым полем другой моды. Связь может быть линейной и нелинейной. Если она окажется достаточно большой, то хаотическая динамика может стать регулярной динамикой.

Следует обратить внимание на тот факт, что коэффициент связи в рассмотренном случае, оказывается значительно большим, чем это было в предыдущих случаях ($\mu = 6$). Возникает вопрос: «Какой величины необходимо выбирать коэффициент связи для подавления режимов с динамическим хаосом?» Дело в том, что в каждой точке фазового пространства скорость разбегания фазовых траекторий различна (различны максимальные показатели Ляпунова). Анализировать все такие точки, в общем случае, малопродуктивно. Имеет смысл, по-видимому, ориентироваться на максимальные и минимальные значения показателей Ляпунова. В качестве примера найдем максимальные показания Ляпунова для невозмущенной системы Лоренца. Уравнение для этих показателей имеет вид:

$$\lambda^3 + \lambda^2(1+b) - \lambda\left[(b+x_0^2) + \sigma(r-z_0) - y_0\right] + \sigma\left[(r-z_0)(b-x_0) - y_0 - x_0 \cdot y_0\right] = 0.$$

Здесь $\{x_0, y_0, z_0\}$ - координаты точки фазового пространства, в которых определяются показатели Ляпунова.

Анализ этого уравнения показывает, что в окрестности нулевой стационарной $\{0;0;0\}$ точки максимальный показатель Ляпунова порядка 13. В окрестности стационарной точки $\{\sqrt{r-1}; \sqrt{r-1}; r-1\}$ он порядка 4.8, а в окрестности стационарной точки $\{-\sqrt{r-1}; -\sqrt{r-1}; r-1\}$ он порядка 6.6. Максимальные показатели Ляпунова при больших значениях фазовых координат $\{20;20;20\}$ порядка 21. Таким образом, мы видим, что в большинстве анализируемых точках показатели Ляпунова достаточно большие. Однако для подавления режима с динамическим хаосом оказалось достаточным ввести параметр связи равный 6 ($\mu = 6$).

**ЭКСПЕРИМЕНТАЛЬНЫЕ ИССЛЕДОВАНИЯ**

В настоящее время имеется значительное количество экспериментальных доказательств существования квантового эффекта Зенона [19-22]. Кратко изложим основные идеи этих экспериментов и некоторые их результаты. В основном, рассматривается трехуровневая система. В большинстве случаев расстояние между нижним (первым) и вторым энергетическими уровнями находится в высокочастотном диапазоне (радиодиапазоне). Энергия третьего уровня значительно больше энергии нижних двух уровней. Переходы между третьим и вторым уровнями запрещены. В подавляющем большинстве случаев, переходы между первым и третьим уровнями находятся в оптическом диапазоне. Переходы же между первым и вторым уровнями находятся в высокочастотных диапазонах. В начальный момент времени населен только нижний основной уровень. На рассматриваемую квантовую систему начинают действовать ВЧ-излучением, частота которого соответствует переходам между первым и вторым уровнями. Характерное время перехода системы из первого на второй энергетический уровень в экспериментах стараются сделать как можно большим. В частности, в работе [22] между первым и вторым уровнями находится виртуальный уровень, а переходы между этими уровнями осуществляются воздействиями двумя высокочастотными импульсами. Через короткое время после воздействия высокочастотным сигналом система подвергается коротким лазерным $\pi$- импульсом. Такой импульс быстро переводит систему с первого энергетического уровня на высокий третий энергетический уровень. Время жизни на этом уровне составляет очень маленькую величину. В большинстве случаев этой величиной можно пренебречь. Поэтому система с третьего уровня практически мгновенно возвращается назад на первый исходный уровень. Предполагается, что такое воздействие лазерным импульсом представляет собой процесс измерения состояния изучаемой системы. Если оказывается, что за время перехода под воздействием высокочастотного сигнала с первого на второй уровень таких лазерных импульсов будет много, то населенность первого уровня оказывается практически неизменной. Такой сценарий проведения экспериментов, в которых был впервые обнаружен квантовый эффект Зенона, в других экспериментах отличается только несущественными деталями. Во всех случаях при обсуждении возможностей использования и реализации квантового эффекта Зенона делалось утверждение, что такие процессы не могут быть реализованы на ядерном уровне, например, на подавление или (ускорение) процессов радиоактивного распада.

В ННЦ "ХФТИ" были проведены две серии экспериментов, результаты которых могут быть объяснены на основе принципа юлы [23-24] (смотри также [5,8]). В первой серии экспериментов [23] в высокочастотном поле находился фотокатод ФЭУ. Было обнаружено достоверное, хорошо наблюдаемое уменьшение фототока при воздействии на фотокатод высокочастотного поля. Зависимость уменьшения фототока от мощности высокочастотного поля находится в хорошем качественном согласии с теоретическими предсказаниями. Причем, чем большей была мощность высокочастотного поля, тем меньшим был фототок. Этот эксперимент интересен тем, что воздействие оказывалось не на состояния электронов в потенциальной яме фотокатода, а на те состояния, на которые должен был перейти электрон в результате поглощения оптического кванта. Следует, однако, сказать, что в настоящее время нет строгого доказательства, что такое уменьшение фототока полностью связано с принципом юлы. Действительно, трудно оценить всю динамику электронов, которые движутся в высокочастотном поле и которые бомбардируют электроды ФЭУ.

Вторая серия экспериментов [24] относится к ядерным процессам. Суть экспериментов, которые были проведены в ННЦ "ХФТИ" и в Киевском институте ядерных исследований, заключалась в следующем. Образец материала, который содержал радиоактивный изомер гафния-178, воздействовали потоками электронов с энергиями от 10 КэВ до 50 КэВ. На образце, как на мишени, энергия электронов конвертировалась в энергию рентгеновского излучения. По предположению несколько выше основного метастабильного состояния гафния-178 должен находиться короткоживущий триггерный уровень. Цель экспериментов заключалась в том, чтобы перевести метастабильное состояние гафния-178 на неустойчивый короткоживущий триггерный уровень. При успехе эксперимента в руках исследователей появляется идеальное ядерное топливо. Действительно, во многих экспериментах наблюдалось различимое ускорение процесса радиоактивного распада под воздействием потока электронов на мишень, содержащую радиоактивный изомер. Однако чаще авторы наблюдали противоположный эффект, когда процесс радиоактивного распада тормозился при воздействии пучка на мишень. Результаты этой серии экспериментов могут быть объяснены сформулированным выше механизмом стабилизации возбужденных состояний. Действительно, в рассматриваемой ядерной системе кроме триггерного уровня имеется большое количество других долгоживущих энергетических состояний. При воздействии пучка электронов на мишень возбуждается широкий спектр рентгеновского излучения. Такое излучение будет переводить ядерную систему не только на определенный энергетический (желательно триггерный) уровень, но и на многие другие долгоживущие уровни. Систему уравнений, которая будет описывать такие переходы, можно представить в виде:

$$i\dot{A}_0 = A_1, \quad i\dot{A}_1 = A_0 + \sum_{k=2}^{N} \mu_k A_k, \quad i\dot{A}_2 = \mu_2 A_1, \quad i\dot{A}_3 = \mu_3 A_1, \ldots i\dot{A}_N = \mu_N A_1. \tag{42}$$

Эта система уравнений описывает связь и переходы рассматриваемой системы с первого ($A_1$) метастабильного уровня на основной уровень ($A_0$) за счет спонтанных переходов и на другие уровни, близкорасположенные к метастабильному уровню. Последние переходы являются индуцированными. Система (42) имеет аналитическое решение:

$$A_1 = \cos(\Omega \cdot t), \quad A_0 = \frac{1}{i \cdot \Omega} \sin(\Omega \cdot t), \quad A_k = \frac{\mu_k}{i \cdot \Omega} \sin(\Omega \cdot t), \quad k = 2, 3, \ldots N,$$

где $\Omega = \sqrt{1 + \sum_{k=2}^{N} \mu_k^2}$.

Из вида этих решений следует, что чем большим будет число энергетических уровней, на которые возможен индуцированный переход, тем меньшей будет вероятность спонтанного распада и меньшей будет вероятность перехода системы на триггерный короткоживущий уровень.

## ЗАКЛЮЧЕНИЕ. ОБСУЖДЕНИЕ РЕЗУЛЬТАТОВ

В заключение сформулируем и обсудим наиболее важные результаты, которые следует из проведенного выше сравнительного анализа механизмов стабилизации.

1. Полученные выше результаты демонстрируют тот факт, что имеется достаточно эффективный и простой механизм, который позволяет стабилизировать неустойчивые состояния. Этим механизмом является принцип юлы. Он прост, нагляден, достаточно универсален – он может быть использован как для стабилизации нестабильных квантовых систем, так и для стабилизации неустойчивых классических систем. Более того, этот принцип дает не только качественные рекомендации для стабилизации, но во многих случаях указывает и на необходимые количественные характеристики. Важным также является тот факт, что стабилизация на основе использования принципа юлы годится не только для систем с замедленным законом распада (неэкспоненциальным законом распада), но и для неустойчивых систем, для которых характерным является привычный экспоненциальный закон распада и даже для стабилизации взрывных неустойчивостей. Относительно последних - следует заметить, что в некоторых случаях численные расчеты показывают, что полной стабилизации не наступает. Происходит только задержка развития процесса неустойчивости. Иногда значительная задержка. Однако этот вопрос в настоящее время находится в состоянии изучения.

2. Как мы видели выше, стабилизация квантовых систем, прежде всего, обусловлена тем фактом, что процесс распада квантовых систем на малых интервалах времени неэкспоненциален. По-видимому, впервые этот результат был получен в работе [1]. Однако до настоящего времени автор не знает случая, когда при анализе квантового эффекта Зенона ссылались на эту работу. Второй фундаментальный процесс, который лежит в основе эффекта Зенона это процесс измерения. Описание процесса измерения в квантовой механике выходит за рамки традиционной квантовой механики. До настоящего времени нет единого мнения среди специалистов, которые занимаются основами квантовой механики, какую роль играет существующий математический аппарат, описывающий процесс измерения. Ярким примером такого неоднозначного мнения являются результаты опроса, который был проведен среди участников конференции "Quantum Physics and the Nature of Reality", которая проходила в июле 2011 в Австрии [25]. Организаторы опроса задавали вопрос: "Как Вы относитесь к существующему состоянию теории, которая описывает процесс измерения?" Ответы были самые разнообразные. От полного одобрения существующей теории до полного отрицания ее полезности.

3. Следует отметить, что многие авторы указывали на тот факт, что существующие экспериментальные результаты, которые подтверждают результаты теории квантового эффекта Зенона, могут быть объяснены в рамках традиционной квантовой механики. Так, в работе Пригожина и соавторов [26] авторам удалось сформулировать в рамках традиционной квантовой механики свойства проекционного оператора, которые оказались достаточными, чтобы объяснить основные экспериментальные результаты, полученные к тому времени относительно квантового эффекта Зенона. Как мы видели выше, стабилизация квантовых систем может вполне быть описана в рамках традиционной квантовой механики.

4. Следует отметить тот факт, что в известных экспериментальных исследованиях эффекта Зенона стабилизация осуществлялась лазерным излучением квантовых переходов, которые располагались в ВЧ-диапазоне (радиодиапазоне), т.е. осуществлялась стабилизация низкочастотных переходов с помощью излучения высоких частот. Важным результатом, полученным выше, является тот факт, что возможна наоборот стабилизация высокочастотных переходов (лазерных, рентгеновских и т.д.) с помощью воздействия низкочастотных полей. Необходимым условием для такой стабилизации, чтобы частота Раби низкочастотных переходов была больше, а лучше значительно больше величины, которая равна обратному времени жизни возбужденных состояний. Это означает, что число низкочастотных квантов должно быть достаточно велико.

5. На наш взгляд, очень важным и необычным результатом является результат, представленный формулами (17a, 17b). Смотрите также рисунки Рис.1b и Рис.1c. Важность его заключается в том, что если в квантовом эффекте Зенона мы должны измерять (воздействовать) то состояние, в котором находится квантовая система, то в общем случае стабилизация может быть реализована путем воздействия и на те состояния, в которые система должна перейти. Таким образом, стабилизацию квантовой системы можно осуществить либо с помощью превращения стационарного неустойчивого состояния в устойчивое динамическое состояние, либо делать динамическими те состояния, в которые система должна перейти. Этот результат особенно важен для стабилизации таких состояний, на которые внешними силами воздействовать либо затруднительно, либо просто невозможно. Примерами могут служить фотоэффект и ядерный бета-распад.

6. Стабилизация классических систем с использованием принципа юлы также оказалась достаточно эффективной. Выяснилось, что имеется разница между механизмом стабилизации неустойчивых классических систем путем воздействия на них или на их параметры внешнего высокочастотного поля и стабилизацией на основе принципа юлы. Главным различием оказался тот факт, что при стабилизации системы в быстроосциллирующем поле стабилизирующее воздействие оказывается не самосогласованным. Динамика стабилизирующей силы не зависит от динамики стабилизируемой системы. При использовании принципа юлы стабилизирующее воздействие является самосогласованным. Его динамика существенно зависит от динамики стабилизируемой системы. Их динамики самосогласованы. В результате в тех случаях, когда для стабилизации изучаемой системы можно использовать принцип юлы, он оказывается более эффективным, чем стабилизация в быстроосциллирующем поле. Этот результат был продемонстрирован на примере стабилизации верхнего (неустойчивого) положения математического маятника (маятника Капицы). Однако очевидно, что этот результат справедлив и в других случаях. Нужно, однако, отметить, что механизм стабилизации с помощью внешнего быстроосциллирующего поля в некотором смысле более универсален. Он проще при реализации и, похоже, может быть использован для более широкого круга физических систем, чем принцип юлы. Действительно, достаточно трудно представить себе физическую реализацию принципа юлы для конкретного математического маятника. В то время как физическая модель маятника Капицы эффективно работает.

7. Следует обратить внимание также на тот факт, что подавление некоторых неустойчивостей может быть объяснено не только механизмом юлы, но и другими механизмами. Так, если рассматривать подавление плазменно-пучковой неустойчивости путем наложения дополнительной связи возбуждаемой плазменной волны с волной внешней электродинамической структуры (см. выше формулы (38)-(39)), то срыв неустойчивости может быть объяснен тем фактом, что такое взаимодействие приводит к расщеплению дисперсионной ветви плазменной волны. В той области, где ветви плазменной волны и волны внешней электродинамической структуры пересекаются, возникает щель, ширина которой пропорциональна коэффициенту связи между этими волнами. В результате, пучковая мода не пересекает ветвь плазменной волны. Это означает, что исчезают условия для черенковского механизма излучения частиц пучка в плазме. Это, в свою очередь, приводит к нарушению условий, необходимых для развития плазменно-пучковой неустойчивости [27]. Аналогичное объяснение можно привести и для процесса подавления распадной неустойчивости. В этом случае связь одной из волн, участвующих в трехволновом взаимодействии, нарушает условие синхронизма между ними. Это нарушение синхронизма приводит к срыву распадных неустойчивостей. Эти примеры можно продолжить. Привлекательность принципа юлы заключается в том, что он объединяет все эти, на первый взгляд, разрозненные процессы. Он позволяет подойти ко всем этим процессам с единой точки зрения.

8. Говоря о подавлении режимов с динамическим хаосом, следует обратить внимание на трудности, которые могут встретиться при определении конкретных физических механизмов, которые будут эквивалентны механизму вращения юлы. Действительно, если для системы Лоренца в том случае, когда она описывает динамику лазерного поля, найти такой механизм оказалось достаточно просто, то для случая, например, движения частиц в потенциале Хенона-Хейлиса это сделать трудно. Действительно, не составляет труда формально использовать принцип юлы, и сорвать режим динамического хаоса для частиц, которые движутся в кристалле (в потенциале Хенона-Хейлиса [28,29]). Однако, как только возникает вопрос, что представляет собой физическое содержание дополнительной стабилизирующей степени свободы, то здесь возникают трудности. Сделаем еще одно замечание. Очень часто режимы с динамическим хаосом анализируются с помощью сечения Пуанкаре. Однако такой способ не всегда дает правильный результат. Действительно, если посмотреть на сечение Пуанкаре для невозмущенной системы Лоренца и для возмущенной, то мы увидим, что произошло уменьшение фазового объема этих точек (в стабилизируемой системе). Однако утверждать, что динамика стала регулярной по этой картине трудно. По этой причине мы не привели ни вида фазового портрета, ни вида сечений Пуанкаре. Они могут указать только на тенденцию к регуляризации.

## СПИСОК ЛИТЕРАТУРЫ